\documentclass[iop,twocolappendix]{emulateapj}

\slugcomment{Accepted for publication in The Astrophysical Journal}

\shorttitle{CEN X-4}
\shortauthors{CHAKRABARTY ET AL.}

\begin{document}

\title{A Hard X-ray Power-Law Spectral Cutoff in Centaurus~X-4}

\author{Deepto Chakrabarty$^1$,
  John~A.~Tomsick$^2$,
  Brian~W.~Grefenstette$^3$,
  Dimitrios~Psaltis$^4$,
  Matteo~Bachetti$^{5,6}$,
  Didier~Barret$^{5,6}$,
  Steven~E.~Boggs$^2$,
  Finn~E.~Christensen$^7$,
  William~W.~Craig$^{8,2}$,
  Felix~F\"urst$^3$,
  Charles~J.~Hailey$^9$,
  Fiona~A.~Harrison$^3$,
  Victoria~M.~Kaspi$^{10}$,
  Jon~M.~Miller$^{11}$,
  Michael~A.~Nowak$^1$,
  Vikram~Rana$^3$,
  Daniel~Stern$^{12}$,
  Daniel~R.~Wik$^{13}$,
  J\"orn~Wilms$^{14}$,
  William~W.~Zhang$^{13}$
}

\affil{$^1$MIT Kavli Institute for Astrophysics and Space Research,
  Massachusetts Institute of Technology, Cambridge, MA 02139, USA\\
$^2$Space Sciences Laboratory, University of California, Berkeley, CA
  94720, USA\\ 
$^3$Cahill Center for Astronomy and Astrophysics, California Institute
  of Technology, Pasadena, CA 91125, USA\\
$^4$Department of Astronomy, University of Arizona, Tucson, AZ 85721,
  USA\\
$^5$Observatoire Midi-Pyr\'{e}n\'{e}es, Universit\'{e} de Toulouse III
  -- Paul Sabatier, 31400 Toulouse, France\\
$^6$CNRS, Institut de Recherche en Astrophysique et Planetologie,
  31028 Toulouse, France\\
$^7$Division of Astrophysics, National Space Institute, Technical
  University of Denmark, 2800 Lyngby, Denmark\\
$^8$Lawrence Livermore National Laboratory, Livermore, CA 94550, USA\\
$^9$Columbia Astrophysics Laboratory, Columbia University, New York,
  NY 10027, USA\\
$^{10}$Department of Physics, McGill University, Montreal, PQ, H3A
  2T8, Canada\\ 
$^{11}$Department of Astronomy, University of Michigan, Ann Arbor, MI
  48109, USA\\
$^{12}$Jet Propulsion Laboratory, California Institute of Technology,
  Pasadena, CA 91109, USA\\
$^{13}$Astrophysics Science Division, NASA Goddard Space Flight
  Center, Greenbelt, MD 20771, USA\\
$^{14}$Dr.\,Karl-Remeis-Sternwarte and Erlangen Centre for
  Astroparticle Physics, Universit\"at Erlangen-N\"urnberg, D-96049
  Bamberg, Germany}

\begin{abstract}
The low-mass X-ray binary Cen X-4 is the brightest and closest ($<$1.2
kpc) quiescent neutron star transient. Previous 0.5--10~keV X-ray
observations of Cen X-4 in quiescence identified two spectral
components: soft thermal emission from the neutron star atmosphere and
a hard power-law tail of unknown origin. We report here on a
simultaneous observation of Cen X-4 with {\em NuSTAR} (3--79 keV) and
{\em XMM-Newton} (0.3--10 keV) in 2013 January, providing the first
sensitive hard X-ray spectrum of a quiescent neutron star transient.
The 0.3--79~keV luminosity was $1.1\times 10^{33}\,D^2_{\rm
  kpc}$~erg~s$^{-1}$, with $\simeq$60\% in the thermal component. We
clearly detect a cutoff of the hard spectral tail above 10 keV, the
first time such a feature has been detected in this source class.  We
show that thermal Comptonization and synchrotron shock origins for the
hard X-ray emission are ruled out on physical grounds. However, the
hard X-ray spectrum is well fit by a thermal bremsstrahlung model with
$kT_e$=18~keV, which can be understood as arising either in a hot
layer above the neutron star atmosphere or in a
radiatively-inefficient accretion flow (RIAF). The power-law cutoff
energy may be set by the degree of Compton cooling of the
bremsstrahlung electrons by thermal seed photons from the neutron star
surface.  Lower thermal luminosities should lead to higher (possibly
undetectable) cutoff energies.  We compare Cen~X-4's behavior with the
PSR~J1023+0038, IGR~J18245$-$2452, and XSS~J12270$-$4859, which have
shown transitions between LMXB and radio pulsar modes at a similar
X-ray luminosity.
\end{abstract}


\section{INTRODUCTION}

Low-mass X-ray binaries (LMXBs) consist of a neutron star (NS) or
black hole (BH) accreting from a low-mass ($\lesssim~1~M_{\odot}$)
stellar companion via Roche-lobe overflow. They may be divided into
two categories: persistent accretors with X-ray luminosity $L_{\rm
  x}\gtrsim$10$^{36}$~erg~s$^{-1}$, and transient systems.  Transient
LMXBs undergo recurrent bright ($L_{\rm
  x}\gtrsim$10$^{36}$~erg~s$^{-1}$) outbursts lasting days to weeks
and then return to long intervals of X-ray quiescence ($L_{\rm
  x}\lesssim$10$^{34}$~erg~s$^{-1}$) lasting months to years.  The
long-term average mass accretion rate of the transients is thus
substantially lower than in the persistent systems, owing to their low
duty cycle.  Transient behavior is understood to arise from a thermal
instability in the outer accretion disk wherein the viscosity (and
thus the mass accretion rate $\dot M$ through the disk) jumps to a
higher value when a critical surface density is reached as the disk
fills up \citep[see][and references therein]{las01}.  The persistent
LMXBs avoid this instability because their higher accretion rates lead
to increased X-ray heating, keeping the disks permanently ionized
\citep{van96,kkb96}. 

For the NS systems, the 0.5--10 keV X-ray spectrum of quiescent LMXB
transients typically consists of two components: a low-energy
(``soft'') $\sim$0.1~keV thermal component, and a high-energy
(``hard'') power-law component with photon index $1<\Gamma<2$, where
photon flux $dN/dE\propto E^{-\Gamma}$.  The soft component is
generally well fit by a hydrogen atmosphere model for the NS.
The leading explanation for the energy source of the soft component is
a deep crustal heating model \citep{bbr98} in which the emission is
powered by heat injected into the NS crust by pycnonuclear reactions
driven by accretion during transient outbursts.  In this model, the
contribution of quiescent accretion is negligible. X-ray spectroscopy of
soft thermal emission in quiescent NS transients has been used to
infer NS radii \citep{bbr98,rbb+99,gsw+13} and to study the thermal
relaxation of NS crusts \citep[see][and references therein]{wdp13}.
However, a possible problem for such studies is that accretion may not have
completely shut off during quiescence, as suggested by the detection
of quiescent variability in the two brightest quiescent NS/LMXBs, Aql X-1
\citep{rbb+02} and Cen X-4 \citep{cis+04,bcb+13}.  There has been
considerable debate as to whether this variability is primarily in the
soft thermal component, the hard power-law component, or both
\citep[e.g.,][]{rbb+02,cs03,cwh+05}.

The origin of the hard power-law tail is unclear.  It is not predicted
by the deep crustal heating model \citep{bbr98}.  Two explanations
have been discussed: synchrotron shock emission from a radio pulsar
wind and Comptonization of the soft thermal photons by a hot corona
\citep{ccm+98}. The synchrotron model is of particular interest given
the recent confirmation that NS/LMXBs can turn on as radio pulsars at
low accretion rates \citep{asr+09,pfb+13}. A difficulty in
discriminating between different models has been the absence of
knowledge about how high the power-law component extends in energy,
owing to lack of sufficient observational sensitivity above 10~keV.
The recent launch of the {\em NuSTAR} hard X-ray telescope provides
the first opportunity to explore this question.

The ideal target with which to address this is Cen X-4 = X1455$-$314
(Galactic coordinates $l=332.2^\circ$, $b=23.9^\circ$), the brightest
quiescent NS/LMXB.  It was discovered in 1969 in the 3--12~keV band
with the {\em Vela 5A/5B} satellites during an extremely bright
($\sim$20 Crab at peak) X-ray outburst lasting over two months
\citep{ceb69,ebc70}.  A second bright ($\sim$4 Crab at peak) X-ray
outburst was detected in 1979 \citep{khs80} along with counterparts in
the optical \citep{cmg80} and radio \citep{hje79,hch+88}, but the
source has been in X-ray quiescence ($\sim 10^5 \times$ fainter) ever
since.

Bright X-ray flashes, now understood as thermonuclear (type I) X-ray
bursts, were observed around the time of both the 1969 and 1979
outbursts \citep{bce72,mik+80}, conclusively establishing the source
as a NS and setting an upper limit on the distance of 1.2$\pm$0.3~kpc
\citep{civ+89}.  A third burst may have been observed by the {\em
  Apollo 15} lunar mission in 1971 \citep{kil09}.  The presence of
thermonuclear bursts indicates that the surface dipole magnetic field
is weak, with $B_{\rm surf}\lesssim 10^{10}$~G \citep{jl80} and most
likely $\sim 10^8$~G (by analogy with other type I bursters).  Optical
photometry and spectroscopy indicate that the binary companion
V822~Cen is a K3-7 V dwarf, the binary period is 15.1~hr, and the
binary mass ratio is $q=0.1755$
\citep{civ+89,tcm+02,dcc+05,swd14}. The best-fit NS mass is
$1.94^{+0.37}_{-0.85} M_\odot$ \citep{swd14}. Given the proximity and
high Galactic latitude of the source, it has extremely low
interstellar extinction and absorption, allowing more sensitive
observations in the ultraviolet and soft X-ray bands than usually
possible for LMXBs \citep{brd+84,mr00,pg11,cbd+13}.  The integrated
values along this line of sight are $A_V=0.362$ \citep{sf11} and
$N_{\rm H}\approx 9\times 10^{20}$~cm$^{-2}$ \citep{dl90,kbh+05}.

Cen X-4 has been observed extensively during X-ray quiescence since
the 1979 outburst, with deep X-ray spectra in the 0.5--10~keV range
previously obtained on six occasions since 1994 \citep[see][and
  references therein; see also \S5.2]{cbm+10}.  A long-term daily
monitoring campaign with {\em Swift} recently demonstrated that the
thermal and power-law components vary together on time scales from
days to months, with no spectral change observed and each component
contributing roughly half the flux \citep{bcb+13}. These authors
concluded that the quiescent X-ray emission in Cen X-4 is primarily
generated by accretion.

In this paper, we present the first sensitive hard X-ray observation
of Cen X-4 in quiescence with {\em NuSTAR}, obtained simultaneously
with a deep {\em XMM-Newton} soft X-ray observation.  We describe the
observations in \S2 and our spectral analysis and results in \S3. We
interpret our results in \S4 and discuss their implications in \S5. 

\section{OBSERVATIONS}

\subsection{NuSTAR}
 
{\em NuSTAR}, the first focusing hard X-ray observatory in orbit, was
launched in 2012 and operates in the 3--79~keV range \citep{har13}. It
consists of two co-aligned telescopes, and the two focal planes (FPMA
and FPMB) are each covered by a 2$\times$2 array of
cadmium-zinc-telluride (CZT) pixel detectors.  Our {\em NuSTAR}
observation of Cen X-4 (ObsID 30001004002) began on 2013 January 20,
20:20 UT and had an elapsed duration of 219 ks, with an on-source
exposure time of 114~ks. The source was imaged on detector~0 in both
FPMA and FPMB.  The data were processed and screened using the
standard pipeline for on-axis point sources in the {\em NuSTAR} Data
Analysis System ({\tt nustardas}) version 1.2.0, along with the {\em
  NuSTAR} calibration database (CALDB) version 20130509.  Light curves
and spectra from both FPMA and FPMB were extracted from a circular
region centered on the source position with a radius of 75 arcsec.

A detailed background model for the source position in each of the two
telescopes was derived by using the {\tt nuskybgd} tool \citep{wik14}
to fit blank sky regions covering the entire field of view for each
focal plane. For faint point sources like Cen X-4, this is more
accurate than the usual method of simply scaling from the background
of a nearby blank sky region because it correctly accounts for the
highly non-uniform background gradients across the detectors.  In both
telescopes, the background was brighter than the source above around
20~keV. The background-subtracted count rates in FPMA and FPMB were
$(4.51\pm 0.09)\times 10^{-2}$ count~s$^{-1}$ and $(4.08\pm
0.09)\times 10^{-2}$ count~s$^{-1}$, respectively. The {\em NuSTAR}
spectra were rebinned so that all but the highest energy bin had a
background-subtracted significance of at least 10$\sigma$.  The
highest energy bin had 5.2$\sigma$ significance in FPMA (20--79~keV)
and 4.3$\sigma$ significance in FPMB (17--79~keV), demonstrating that the
source was detected beyond 20~keV.

\subsection{XMM-Newton}

The {\em XMM-Newton} observatory, launched in 1999, is a focusing
X-ray telescope operating in the 0.3--12~keV range \citep{jan01}.  Our
{\em XMM-Newton} observation of Cen X-4 (ObsID 0692790201) began on
2013 January 21, 13:01 UT and had a duration of 35~ks; this was
simultaneous with part of our {\em NuSTAR} observation.  We used the
data from all three co-aligned imaging X-ray cameras \cite[EPIC-pn,
  EPIC-MOS1, and EPIC-MOS2;][]{str01,tur01}.  The cameras were
operated in full-frame mode with the thin optical-blocking filter in
place, resulting in a time resolution of 73~ms for EPIC-pn and 2.6~s
for the MOS cameras. The data were reduced using the {\em XMM-Newton}
Scientific Analysis System (SAS) v13.0.1, along with the latest
calibration files available as of 2013 July 13. We reprocessed the
data using {\ttfamily epproc} and {\ttfamily emproc} to produce new
event files and applied standard event filtering.  We identified
background flare intervals by constructing a light curve of the
10--12~keV EPIC-pn data using events from the entire field of view,
and searching for intervals where the count rate exceeded
1~count~s$^{-1}$.  We found three short flares; these intervals were
removed from our Cen X-4 event lists.  The net exposure time was
27\,ks for EPIC-pn and 30\,ks for each of the EPIC-MOS units.

For all three detectors, we initially extracted light curves and
spectra from a circular region centered on the source position with a
radius of 43.5 arcsec.  For the MOS detectors, we measured the
background using a square blank-sky region 3$\times$3 arcmin in
size. For the EPIC-pn detector, an important consideration is that the
outer parts of the field-of-view include photons due to Cu
fluorescence in the instrument while the inner parts do not.  As
Cen~X-4 was in the region without the Cu emission line, we chose a
rectangular region near Cen~X-4 for determining the EPIC-pn
background.  The background-subtracted count rates in the 0.3--10~keV
range were 2.58$\pm$0.01\,count~s$^{-1}$ in EPIC-pn,
0.640$\pm$0.005\,count~s$^{-1}$ in EPIC-MOS1, and
0.619$\pm$0.005\,count~s$^{-1}$ in EPIC-MOS2.

These count rates are higher than the those observed in previous {\em
  XMM-Newton} observations \citep{cbm+10}.  In fact, the count rates
in both the EPIC-MOS and EPIC-pn detectors are close to the threshold
where photon pileup effects may begin to distort the measured
spectra, particularly during the flares\footnote{See Table 3 in
  \S3.3.2 of the {\em XMM-Newton Users Handbook}, v2.11, 2013, {\tt
    http://heasarc.gsfc.nasa.gov/docs/xmm/uhb/}.}.  Moreover, our
preliminary analysis found that the two EPIC-MOS spectra each have
significant systematic differences with the EPIC-pn spectrum above
2~keV. As a precaution, we reextracted both the MOS and pn data from
an annular region centered on the source position with an outer radius
of 43.5 arcsec and an inner radius of 10 arcsec, thus excluding the
core of the point-spread function (PSF) where any pile-up would occur
(at the expense of reduced counting statistics). The MOS and pn
spectra from the annular region are mutually consistent.  This annulus-only
data set still includes sufficient counts to obtain a good measurement
of the soft X-ray spectrum. 
\vspace*{0.3in}

\section{DATA ANALYSIS AND RESULTS}

\subsection{Timing}

\begin{figure}[t]
\begin{center}
\includegraphics[width=0.47\textwidth]{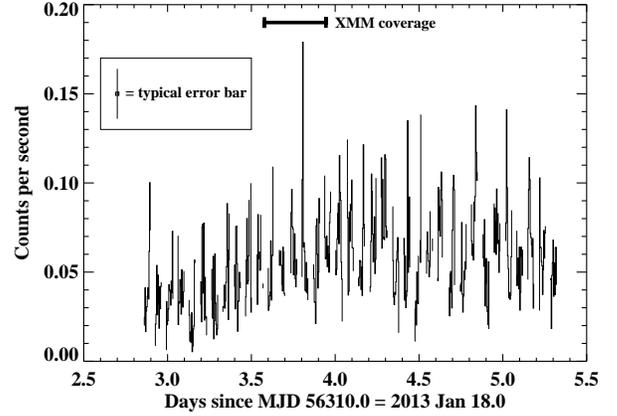}
\end{center}
\caption{Background-subtracted 3--20 keV light curve of Cen X-4
  measured with {\em NuSTAR}, binned at 300~s resolution. The average
  of the FPMA and FPMB count rates is plotted, and the size of a
  typical error bar is shown.  The data gaps are due to
  Earth occultations. The source intensity varies significantly on time
  scales of minutes to hours. The interval with simultaneous coverage with {\em
    XMM-Newton} is indicated, and includes a bright flare around MJD
  56313.8 (see Figure~\ref{fig_lc_xmm}).  \label{fig_lc}}
\end{figure}

\begin{figure}[t]
\begin{center}
\includegraphics[width=0.47\textwidth]{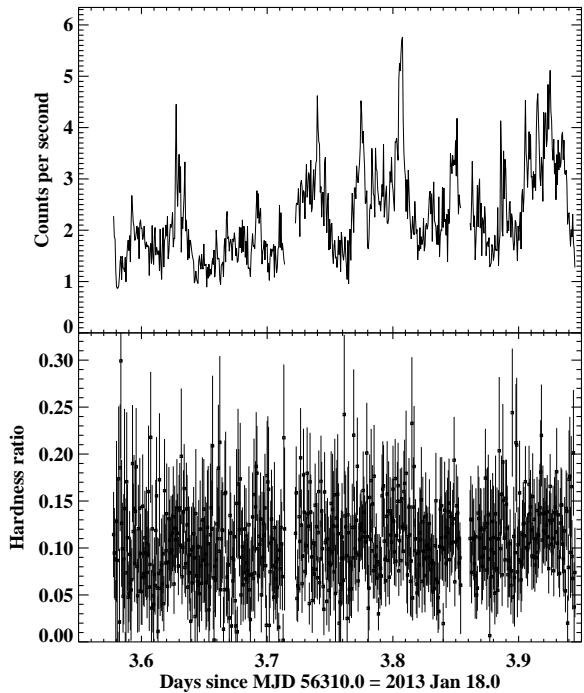}
\end{center}
\caption{{\em Top panel:} Background-subtracted 0.3--10 keV light
  curve of Cen~X-4 measured with {\em XMM-Newton}/EPIC-pn, binned at
  50~s resolution. The typical error bar is $\pm$0.16 counts~s$^{-1}$.
  The short data gaps are intervals that were excluded owing to
  strong background flares. The source intensity shows strong flaring
  behavior, varying by more than a factor of two on time scales as
  short as a few minutes. The brightest flare, around MJD 56313.8, is
  also visible in the {\em NuSTAR} light curve (see
  Figure~\ref{fig_lc}). {\em Bottom panel:} Spectral hardness of the
  count rate shown in the top panel, constructed by taking the ratio
  of the 2--10~keV and 0.3--1.0~keV count rates.  There is no evidence
  for significant spectral changes during the flaring intervals,
  although there is weak evidence for a small change during the flare
  at MJD 56313.63.  \label{fig_lc_xmm}}  
\end{figure}

The light curves for the {\em NuSTAR} and {\em XMM-Newton}
observations are shown in Figure~\ref{fig_lc} and the top panel of
Figure~\ref{fig_lc_xmm}, respectively.  (The {\em XMM-Newton} light
curve uses the full data set, not the annulus-only data.)  Both light
curves vary significantly on time scales of a few minutes.  The {\em
  XMM-Newton} light curve clearly exhibits flaring activity; the
strongest of these flares, around MJD 56313.8, is also easily visible
in the {\em NuSTAR} light curve.  The fractional excess
root-mean-squared (rms) variability\footnote{This is a measure of the
  intrinsic source variability in excess of Poisson counting noise;
  see, e.g., \citet{vew+03}.} $F_{\rm var}$ was 37$\pm$6\% in the
{\em NuSTAR} light curve and 37$\pm$2\% in the {\em
  XMM-Newton}/EPIC-pn light curve.  For comparison, a value of
73.0$\pm$1.5\% was measured in 60 short {\em Swift} observations made
over three months \citep{bcb+13}.

\begin{figure}[t]
\begin{center}
\includegraphics[width=0.47\textwidth]{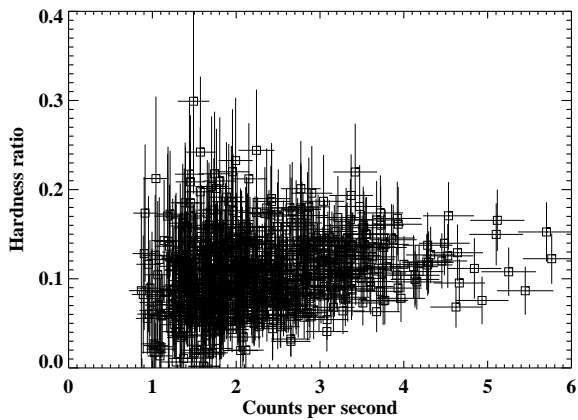}
\end{center}
\caption{Hardness-intensity diagram for {\em XMM-Newton}/EPIC-pn
  observation of Cen X-4. The $x$-axis shows the 0.3--10 keV count
  rate, and the $y$-axis shows the spectral hardness (defined as the
  ratio of the 2--10~keV and 0.3--1.0~keV count rates).  There is no
  evidence for significant spectral changes as a function of source
  intensity. \label{fig_hi}}  
\end{figure}

A comparison of the 0.3--1 keV and 2--10 keV {\em XMM-Newton}/EPIC-pn
light curves shows no evidence for significant spectral changes during
these flares, although there is weak evidence for a small change
during the flare at MJD 56313.63 (Figure~\ref{fig_lc_xmm}, bottom
panel; Figure~\ref{fig_hi}).  A cross-correlation analysis indicates
that the flares in these two bands are simultaneous to within
$\lesssim$30~s. We searched the {\em NuSTAR} light curve for evidence
of orbital variability by folding at the 15.1~hr binary period
\citep{swd14}.  No orbital modulation was detected.

\subsection{Spectroscopy}

Since there is no evidence for significant spectral variability during
the flares (Figure~\ref{fig_lc_xmm}, bottom panel;
Figure~\ref{fig_hi}), we chose to include the flare intervals for our
spectral analysis.  We performed joint spectral fits of the {\em
  NuSTAR} and {\em XMM-Newton} (annulus) data with XSPEC version 12.8
spectral fitting software \citep{arn96} using $\chi^2$-minimization.
To allow for systematic calibration differences between the different
detectors, we included a constant multiplicative factor in the
model. This constant was set to unity for {\em XMM-Newton}/EPIC-pn but
allowed to vary for the other detectors. All other spectral model
parameters were tied together across the three instruments.
Interstellar absorption was modeled using the {\tt tbabs} model
\citep{wam00} along with photoionization cross-sections from
\citet{ver96}.

We tried two different models for the soft thermal component. We first
fit a passively cooling neutron star atmosphere using the {\tt
  nsatmos} model \citep{hrn+06}, which assumes a pure H atmosphere and
a negligible surface magnetic field ($B\lesssim 10^8$~G), and also
includes the effects of surface gravity, heat conduction by electrons,
and self-irradiation.  In applying {\tt nsatmos}, we fixed the source
distance at 1.2~kpc \citep{civ+89} and assumed that the atmospheric
emission was coming from the entire NS surface.  We found that it was
not possible to constrain the NS mass $M$ and radius $R$ when both
were allowed to vary, as a wide range of $M$-$R$ pairs gave acceptable
fits.  We therefore fixed the NS mass to $M=1.9 M_\odot$
\citep{swd14}.  As an alternative model, we also tried fitting to
synthetic spectra of \citet{ztz+95,zct+01} for thermal emission from
unmagnetized NSs with a pure H atmosphere in the presence of very low
accretion rates.  These spectra have been implemented as the XSPEC
additive table model {\tt zamp} and are parametrized in terms of the
observed accretion luminosity scaled to the Eddington rate
($L_\infty/L_E$), with fixed NS mass $M=1.4 M_\odot$ and true
radius $R=12.4$~km. In practice, the model shapes from {\tt nsatmos}
and {\tt zamp} are essentially identical \citep[see, e.g.][]{szz+11},
but it is useful to demonstrate that models that explicitly include
accretion are consistent with the data.

\begin{figure}[t]
\begin{center}
\includegraphics[width=0.47\textwidth]{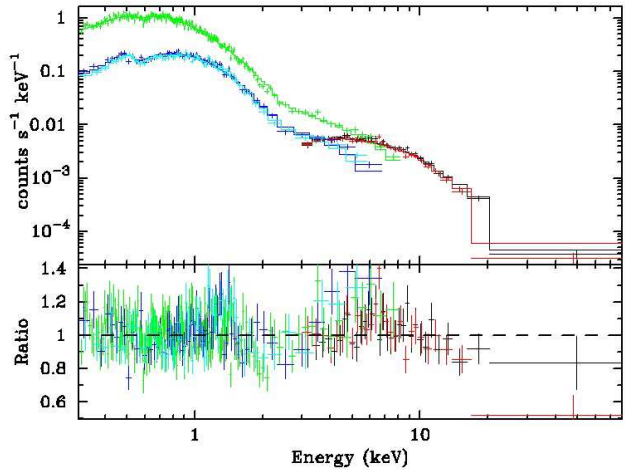}
\end{center}
\caption{Count spectrum and fit for our joint {\em XMM-Newton} and {\em
    NuSTAR} observation of Cen X-4, using an unbroken power-law to
  model the hard X-ray spectrum. The model used is {\tt
    tbabs*(nsatmos+powerlaw)}. The {\em XMM-Newton} data were 
  extracted from an annulus in order to avoid possible photon pileup
  effects (see \S2.2). The green points are the {\em XMM-Newton}
  EPIC-pn data; the dark and light blue points are the {\em
    XMM-Newton} EPIC MOS1 and MOS2 data, respectively; and the black
  and red points are the {\em NuSTAR} FPMA and FPMB data,
  respectively.  The residuals are consistent with a break in the
  power law spectrum around 7 keV or a cutoff in the 10--20 keV
  range. \label{fig_pl}}
\end{figure}

\begin{figure}
\begin{center}
\includegraphics[width=0.47\textwidth]{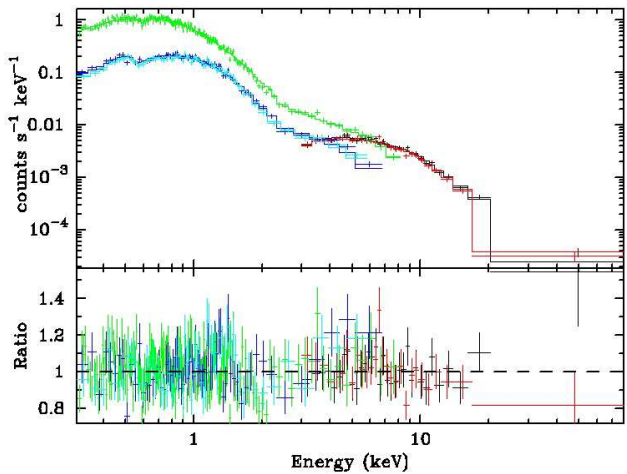}
\end{center}
\caption{Count spectrum and model fit for our joint
  {\em XMM-Newton} and {\em NuSTAR} observation of Cen X-4, with the
  hard X-ray spectral cutoff included in the model. A bremsstrahlung
  model is shown: {\tt tbabs*(nsatmos+bremss)}. The color scheme is
  the same as in Figure~\ref{fig_pl}. The fit parameters are
  shown in Table~1. \label{fig_spec}} 
\end{figure}

\begin{figure}
\begin{center}
\includegraphics[width=0.47\textwidth]{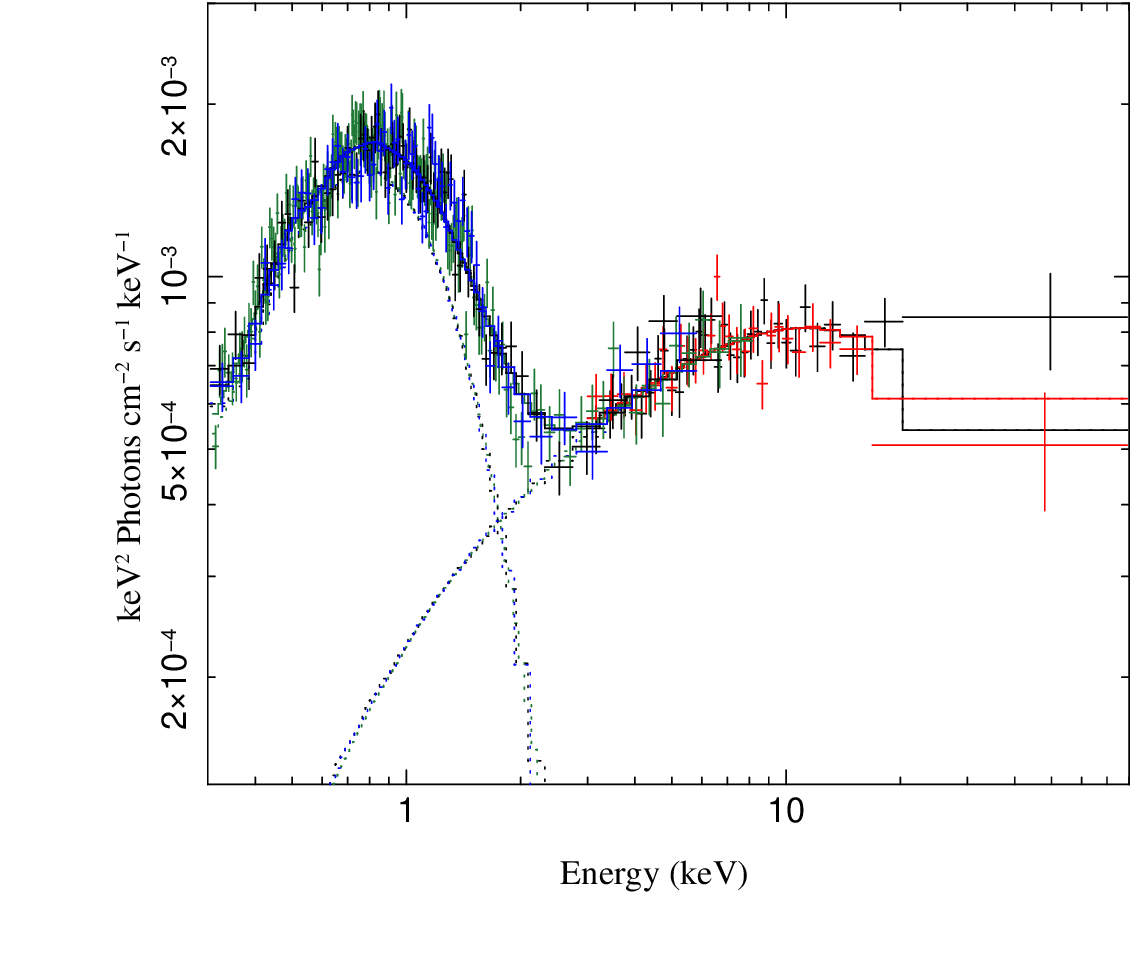}
\end{center}
\caption{The unfolded $\nu  F_\nu$ spectrum of Cen X-4 corresponding
  to the {\tt tbabs*(nsatmos+bremss)} fit shown in
  Figure~\ref{fig_spec}. The individual model components are denoted
  by the dotted lines. The cutoff of the hard spectral component
  above 10~keV is clearly visible.\label{fig_uf}} 
\end{figure}

Although previous 0.5--10~keV X-ray observations of Cen X-4 in
quiescence were all well fit by an absorbed passive NS atmosphere plus
power-law model \citep[see][and references therein]{cbm+10}, the {\tt
  tbabs*(nsatmos+powerlaw)} model does not provide a satisfactory fit
to the high-energy data in our combined {\em NuSTAR}+{\em XMM-Newton}
data covering the 0.3--79~keV range.  The residuals indicate the
presence of a spectral break or cutoff in the 10--20~keV range (see
Figure~\ref{fig_pl}).  We found several different models provided a
good fit for the hard component: a cut-off power-law ({\tt cutoffpl}),
a broken power-law ({\tt bknpower}), thermal Comptonization ({\tt
  comptt}; Titarchuk 1994; Hua \& Titarchuk 1994), and thermal
bremsstrahlung ({\tt bremss}).  In each case, this component was
combined with {\tt tbabs} and either {\tt nsatmos} or {\tt zamp}. In
all cases, the $N_{\rm H}$ value is consistent with the integrated
interstellar value along the line of sight \citep{dl90,kbh+05}.
We found no evidence for fluorescent Fe line emission in the
spectrum.  The upper limit on the equivalent width of a narrow Gaussian
Fe emission line at 6.4~keV is $<$120~eV (90\%-confidence).  

The spectral fits with {\tt nsatmos} are summarized in Table~1; those
with {\tt zamp} are summarized in Table~2.  A typical fit is shown in
Figure~\ref{fig_spec}.  The corresponding unfolded spectrum is shown in
Figure~\ref{fig_uf}. We are able to obtain reasonably good fits with a wide
variety of models. We note that the only spectral shape parameter in
the {\tt zamp} model is the accretion luminosity $L_\infty$ observed
at infinity; one must compare this with the measured flux and source
distance to check for self-consistency. We find that the
best-fit values of $L_\infty$ are roughly consistent with the measured
soft flux for the assumed distance of 1.2~kpc.

Using the {\tt tbabs*(nsatmos+bremss)} model, we find an average
(absorbed) flux of $3.8\times 10^{-12}$~erg~cm$^{-2}$~s$^{-1}$ in the
soft (0.3--3~keV) band and $3.3\times 10^{-12}$~erg~cm$^{-2}$~s$^{-1}$
in the hard (3--79~keV) band. The unabsorbed soft luminosity was
$L_{\rm soft}=6.6\times 10^{32} D^2_{\rm kpc}$ erg~s$^{-1}$
(0.3--3~keV), the unabsorbed hard luminosity was $L_{\rm
  hard}=4.0\times 10^{32} D^2_{\rm kpc}$ erg~s$^{-1}$ (3--79~keV), and
the total unabsorbed luminosity was $L_x=1.1\times 10^{33} D^2_{\rm
  kpc}$ erg~s$^{-1}$ (0.3--79~keV), where $D_{\rm kpc}$ is the source
distance in kiloparsecs.  For comparison with previous observations,
the 0.5--10~keV unabsorbed luminosity was $6.4\times 10^{32} D^2_{\rm
  kpc}$ erg~s$^{-1}$. This is the highest 0.5--10~keV luminosity ever
measured from Cen X-4 in quiescence.  The next highest observation
was 2.6 times fainter in 2001 \citep{cis+04}, and the faintest
observation was 11.5 times fainter in 2009 \citep{cbm+10}. The
thermal (NS atmosphere) contribution to the 0.5--10~keV luminosity on
our observation is 59\%.  This is consistent with previous observations,
where the thermal fraction has always been 50--60\% \citep{cbm+10,bcb+13}.

\begin{deluxetable*}{lcccccccc}
\tablecaption{SPECTRAL FITS WITH A PASSIVE NEUTRON STAR
  ATMOSPHERE\tablenotemark{a}\label{tbl-1}}  
\tablewidth{0pt}
\tablehead{ & & & \multicolumn{6}{c}{Hard spectral component}\\ \cline{4-9}
\colhead{Parameter} & \colhead{Symbol} & \colhead{Units} &  
\colhead{\tt cutoffpl} & \colhead{\tt bknpow} & \colhead{\tt
  comptt(disk)} & \colhead{\tt comptt(sphere)} & \colhead{\tt bremss}  
}
\startdata
\multicolumn{3}{l}{\underline{{\em Absorption and} {\tt nsatmos} {\em
      parameters}}} \\ 
Absorption column  & $N_{\rm H} $ & $10^{21}$ cm$^{-2}$  & 0.88(4)  &
0.89(3) & 0.92(4) & 0.92(4) & 0.87(5) \\
Temperature (unredshifted) & $\log T$ & K & 6.16(4) & 6.18(1) &
6.20(27) & 6.19(23) & 6.20(10) \\ 
Mass (fixed)  & $M$ & $M_\odot$ & 1.9 & 1.9 & 1.9 & 1.9 & 1.9 \\
Radius  & $R$ & km & 9.2(1.0) & 8.6(2.6) & 8.5(5.7) & 8.6(4.9) & 8.1(1.8) \\
Covering fraction (fixed) & $K_{\rm NS}$  &$\cdots$ & 1 & 1 & 1 & 1 & 1 \\
Distance (fixed) & $D$ & kpc & 1.2 & 1.2 & 1.2 & 1.2 & 1.2 \\
\\
\multicolumn{3}{l}{\underline{\em Hard spectral component parameters}} \\
Normalization & $K$ & $10^{-4}$  & 2.3(2)\tablenotemark{b} &
2.2(2)\tablenotemark{b} & 0.7(1)\tablenotemark{b} & 0.7(1)\tablenotemark{b} &
4.1(1)\tablenotemark{c} \\  
Photon index & $\Gamma$ &$\cdots$ & 1.02(10) & 1.26(10) & $\cdots$ & $\cdots$&
$\cdots$\\ 
Cutoff or break energy & $E_c$, $E_b$& keV & 10.4(1.4) & 5.8(3) & $\cdots$
& $\cdots$& $\cdots$\\ 
Photon index 2 & $\Gamma_2$ &$\cdots$ & $\cdots$ &  2.02(5) & $\cdots$
& $\cdots$ & $\cdots$ & \\ 
Electron temperature & $kT_e$ & keV & $\cdots$ & $\cdots$ &
6.4(1.1) & 6.4(1.1) & 18.2(1.0) \\ 
Seed temperature & $kT_0$ & keV & $\cdots$ & $\cdots$ & 0.74(8) &
0.74(7) & $\cdots$ \\ 
Scattering optical depth & $\tau_{\rm es}$ & $\cdots$& $\cdots$ &
$\cdots$ & 4.0(5) & 8.8(1.0) & $\cdots$ \\
\\
\multicolumn{3}{l}{\underline{\em Instrument multiplicative constants}} \\
{\em NuSTAR}/FPMA & $C_{\rm FPMA}$ & $\cdots$& 1.04(3) & 1.03(3) & 1.02(3) &
1.02(3) & 1.09(3) \\ 
{\em NuSTAR}/FPMB & $C_{\rm FPMB}$ &$\cdots$ & 1.07(4) & 1.05(4) & 1.05(4) &
1.05(4) & 1.12(4) \\
{\em XMM}/EPIC-pn (fixed) & $C_{\rm pn}$ &$\cdots$ & 1 & 1 & 1 & 1 & 1 \\
{\em XMM}/EPIC-MOS1 & $C_{\rm MOS1}$ &$\cdots$ & 1.04(1) & 1.04(1) & 1.04(1) &
1.03(1) & 1.04(1) \\
{\em XMM}/EPIC-MOS2 & $C_{\rm MOS2}$ &$\cdots$ & 0.97(1) & 0.97(1) & 0.97(1) &
0.97(1) & 0.97(1) \\
\\
Fit statistic & $\chi^2_\nu$/dof & $\cdots$& 1.155/380 & 1.123/379 &
1.130/379 & 1.129/379 & 1.167/381
\enddata
\tablenotetext{a}{1$\sigma$ uncertainties in last digits shown in
  parentheses.}
\tablenotetext{b}{Flux density at 1 keV in units of photons
  cm$^{-2}$ s$^{-1}$ keV$^{-1}$.}
\tablenotetext{c}{$(3.02\times 10^{-15}/4\pi D^2)\int n_e^2 dV$ in
  units of cm$^{-5}$.}
\end{deluxetable*}

\begin{deluxetable*}{lcccccccc}
\tablecaption{SPECTRAL FITS WITH AN ACCRETING NEUTRON-STAR
  ATMOSPHERE\tablenotemark{a}\label{tbl-2}}   
\tablewidth{0pt}
\tablehead{ & & & \multicolumn{6}{c}{Hard spectral component}\\ \cline{4-9}
\colhead{Parameter} & \colhead{Symbol} & \colhead{Units} &  
\colhead{\tt cutoffpl} & \colhead{\tt bknpow} & \colhead{\tt
  comptt(disk)} & \colhead{\tt comptt(sphere)} & \colhead{\tt bremss}  
}
\startdata
\multicolumn{3}{l}{\underline{{\em Absorption and} {\tt zamp} {\em
      parameters}}} \\ 
Absorption column  & $N_{\rm H} $ & $10^{21}$ cm$^{-2}$  & 0.89(3)  &
0.87(6) & 0.69(5) & 0.69(5) & 0.93(5) \\
Luminosity & $\log L_\infty/L_E$ & $\cdots$ & $-$5.15(3) & $-$5.13(4)
& $-$4.98(3) & $-$4.98(3) & $-$5.19(3) \\
Mass (fixed)  & $M$ & $M_\odot$ & 1.4 & 1.4 & 1.4 & 1.4 & 1.4 \\
Radius (fixed) & $R$ & km & 12.4 & 12.4 & 12.4 & 12.4 & 12.4 \\
Normalization & $K_{\rm zamp}$ & $10^{-3}$ & 1.25(2)\tablenotemark{b}
& 1.25(2)\tablenotemark{b} & 1.39(2)\tablenotemark{b} &
1.39(2)\tablenotemark{b} & 1.23(2)\tablenotemark{b} \\
\\
\multicolumn{3}{l}{\underline{\em Hard spectral component parameters}} \\
Normalization & $K$ & $10^{-4}$  & 2.5(2)\tablenotemark{b} &
2.3(2)\tablenotemark{b} & 0.6(1)\tablenotemark{b} & 0.6(1)\tablenotemark{b} &
4.2(1)\tablenotemark{c} \\  
Photon index & $\Gamma$ &$\cdots$ & 1.08(9) & 1.29(7) & $\cdots$ & $\cdots$&
$\cdots$\\ 
Cutoff or break energy & $E_c$, $E_b$& keV & 11.2(1.5) & 5.9(3) & $\cdots$
& $\cdots$& $\cdots$\\ 
Photon index 2 & $\Gamma_2$ &$\cdots$ & $\cdots$ &  2.03(5) & $\cdots$
& $\cdots$ & $\cdots$ & \\ 
Electron temperature & $kT_e$ & keV & $\cdots$ & $\cdots$ &
7.3(1.6) & 7.6(1.7) & 17.9(1.0) \\ 
Seed temperature & $kT_0$ & keV & $\cdots$ & $\cdots$ & 0.81(4) &
0.81(4) & $\cdots$ \\ 
Scattering optical depth & $\tau_{\rm es}$ &$\cdots$ & $\cdots$ &
$\cdots$ & 3.6(6) & 7.8(1.2) & $\cdots$ \\
\\
\multicolumn{3}{l}{\underline{\em Instrument multiplicative constants}} \\
{\em NuSTAR}/FPMA & $C_{\rm FPMA}$ &$\cdots$ & 1.04(3) & 1.03(3) & 1.01(3) &
1.01(3) & 1.08(3) \\ 
{\em NuSTAR}/FPMB & $C_{\rm FPMB}$ &$\cdots$ & 1.07(3) & 1.05(4) & 1.04(4) &
1.04(4) & 1.10(4) \\
{\em XMM}/EPIC-pn (fixed) & $C_{\rm pn}$ &$\cdots$ & 1 & 1 & 1 & 1 & 1 \\
{\em XMM}/EPIC-MOS1 & $C_{\rm MOS1}$ &$\cdots$ & 1.04(1) & 1.04(1) & 1.03(1) &
1.03(1) & 1.03(1) \\
{\em XMM}/EPIC-MOS2 & $C_{\rm MOS2}$ &$\cdots$ & 0.97(1) & 0.97(1) & 0.97(1) &
0.97(1) & 0.97(1) \\
\\
Fit statistic & $\chi^2_\nu$/dof & $\cdots$& 1.156/380 & 1.129/379 &
1.109/379 & 1.109/379 & 1.162/381
\enddata
\tablenotetext{a}{1$\sigma$ uncertainties in last digits shown in
  parentheses.}
\tablenotetext{b}{Flux density at 1 keV in units of photons
  cm$^{-2}$ s$^{-1}$ keV$^{-1}$.}
\tablenotetext{c}{$(3.02\times 10^{-15}/4\pi D^2)\int n_e^2 dV$ in
  units of cm$^{-5}$.}
\end{deluxetable*}

\section{INTERPRETING THE SPECTRAL CUTOFF}

We now discuss what can be inferred from our observed high-energy
spectral cutoff.  We introduce dimensionless variables for
parametrizing the NS mass $M= 1.9\,M_{1.9} M_\odot$ and radius $R=
10\, R_{10}$~km.  It is also convenient to scale the luminosity and
accretion rate to the Eddington critical values,
\begin{eqnarray}
L_E & = & 2.9\times 10^{38}\,M_{1.9}\left(\frac{1+X}{1.7}\right)^{-1}
    {\rm\ erg\ s}^{-1}  \\
\dot M_E & = & 1.8\times 10^{-8}\,R_{10}
    \left(\frac{1+X}{1.7}\right)^{-1} M_\odot {\rm\ yr}^{-1} \;,
\end{eqnarray}
where $X=0.7$ is the hydrogen mass fraction for cosmic abundances. Our
observed luminosities can then be written as $L_{\rm soft}/L_E =
2.3\times 10^{-6}\,M_{1.9}^{-1}$, $L_{\rm hard}/L_E = 1.4\times
10^{-6}\,M_{1.9}^{-1}$, and $L_x/L_E = 3.8\times 10^{-6}\,M_{1.9}^{-1}$. It is
clear that the quiescent mass accretion rate $\dot M_{\rm NS}$ onto the
NS must also be very low.  We can set an upper limit by assigning all
of the observed X-ray luminosity to accretion, $\dot M_{\rm NS} \leq
L_xR/GM$. We then have $\dot M_{\rm NS}/\dot M_E \leq 3.7\times
10^{-6}\,M_{1.9}^{-1}$.  At such extremely low accretion rates, the
outer accretion disk is likely to transition into a quasi-spherical,
radiatively-inefficient accretion flow (RIAF) at a transition radius
$r_t\sim 10^4 R_{\rm Sch} \sim 6\times 10^9\,M_{1.9}$~cm, where
$R_{\rm Sch}=2GM/c^2 = 5.6\,M_{1.9}$~km is the Schwarzschild radius
\citep{ny95,nar98,men+99}. 

\vspace*{0.3in}

\subsection{Comptonization}
\label{sec-compt}

One way to explain a cutoff power-law spectrum is thermal
Comptonization, where soft thermal seed photons (like those from the
NS atmosphere) are Compton scattered from a corona of hot electrons.
\citet{mm01} have previously argued that Comptonization of the NS
atmosphere photons cannot account for the 1--10~keV power law observed
in Cen X-4 because the observed luminosity in the power-law component
is too high relative to the soft luminosity. We show here that a
thermal Comptonization model for the hard spectral cutoff in Cen~X-4 is
not physically self-consistent, despite providing an acceptable fit to
our observed spectrum. For simplicity, we consider the spherical
geometry case; the results for a disk geometry are qualitatively
similar.

Our {\tt comptt} fits yielded an electron temperature of
$kT_e=6$--8~keV and an electron scattering optical depth of
$\tau_{\rm es}=8$--9 for a spherical geometry (see Tables~1 and
2). This is an unusually low $kT_e$ for a Comptonized plasma,
reflecting the observed spectral cutoff.  It is likewise an unusually
high $\tau_{\rm es}$; this follows from the fact that $kT_e$ and
$\tau_{\rm es}$ are inversely proportional for a fixed power-law index
  $\Gamma$ \citep{tl95}. The optical depth in a corona of radius
$r_c$ is related to the electron density $n_e$ by
\begin{equation}
\tau_{\rm es} = \sigma_T \int_R^{r_c} n_e(r) dr \,,
\end{equation}
where $\sigma_T$ is the Thomson cross-section. As we see below, it is
our high value of $\tau_{\rm es}$ that makes it difficult to find a
physically self-consistent Comptonization model. 

\subsubsection{Comptonization Above the NS Atmosphere or the NS
  Magnetosphere} 

We begin by considering the possibility of a hot, optically thin layer
above the NS atmosphere. Based on our measured $\tau_{\rm es}$, the
electron density in the layer would be $n_e = 1.4\times
10^{22}\,h_3^{-1}$~cm$^{-3}$, where $h= 10^3\,h_3$~cm is the thickness
of the layer.  This relatively dense scattering layer would itself be
a source of significant thermal bremsstrahlung emission, with
$kT_e\simeq 7$~keV. Despite the large scattering optical depth
$\tau_{\rm es}$, the {\em effective} optical depth of the medium at
2~keV (including both scattering and free-free absorption) is only
$\tau_{\rm eff}\simeq \sqrt{\tau_{\rm es}\tau_{\rm ff}}=0.4\,
h_3^{-1/2}$ \citep{rl79}, where $\tau_{\rm ff}=0.02\,h_3^{-1}$ is the
free-free absorption optical depth at 2~keV.  (At higher energies,
$\tau_{\rm ff}$ will be even smaller.) We can thus use the emissivity for
optically thin thermal bremsstrahlung to compute the expected
luminosity from the dense scattering layer, $3.6\times
10^{36}\,h_3^{-1}$~erg~s$^{-1}$ $\approx 10^{-2}\, h_3^{-1}\,L_E$.
This is orders of magnitude brighter than what we observe and 
can be ruled out.

We also consider a corona around the NS magnetosphere.  In X-ray
quiescence, we can scale the magnetospheric radius $r_m$
to the corotation radius (where the Keplerian and stellar angular
velocities are equal), which is given by 
\begin{equation}
r_{\rm co} = \left(\frac{GMP^2}{4\pi^2}\right)^{1/3}
      = 39\,P_{\rm 3ms}^{2/3}\,M_{1.9}^{1/3} \mbox{\rm\ km} \,,
\label{eq_corot}
\end{equation}
where we have written the (unknown) NS spin period as $P=3\,P_{\rm
  3ms}$~ms. We assume a spherically symmetric corona with inner radius
$r_{\rm co}$ and scale size $h\sim r_{\rm co}$. Then, following the
calculations in the previous paragraph, we find an electron
density $n_e\simeq 4\times 10^{18}\, (h/r_{\rm co})^{-1}$~cm$^{-3}$, an
effective optical depth of $\tau_{\rm eff}\simeq\sqrt{\tau_{\rm
    es}\tau_{\rm ff}}\simeq 0.008\, (h/r_{\rm co})^{-1/2}$ at 2~keV,
and a predicted bremsstrahlung luminosity from the scattering corona
of $4\times 10^{35}\,(h/r_{\rm co})^{-1}$~erg~s$^{-1}$ $\approx
10^{-3}\, (h/r_{\rm   co})^{-1}\, L_E$.  This is, again, orders of
magnitude brighter than what we observe and can be ruled out.   

\subsubsection{Comptonization in the Accretion Flow}

We next consider Comptonization in a RIAF-like spherical accretion flow with
a radially uniform mass inflow rate $\dot M$ at infall velocity $v_r =
\eta\sqrt{GM/r}$ (where $\eta \leq 1$). The electron density will
vary with distance $r$ from the NS as
\begin{equation}
n_e(r) = \frac{\dot M_{\rm NS}}{4\pi\eta\mu m_p (GMr^3)^{1/2}}\;,
\label{eq:n_e}
\end{equation}
where $\mu$ is the mean molecular weight and $m_p$ is the proton
mass. We then find that the optical depth cannot exceed 
\begin{equation}
\tau_{\rm es,max} \approx 10^{-3} \left(\frac{\eta}{0.1}\right)^{-1}
        \left(\frac{\dot M_{\rm NS}/\dot M_E}{4\times 10^{-6}}\right),
\label{eq_tau}
\end{equation}
independent of $r_c$.  This is orders of magnitude smaller than our
{\tt comptt} fit values for $\tau_{\rm es}$. 

The only way to obtain higher optical depths is to assume that the
mass inflow rate $\dot M$ varies with $r$ in such a way that only a
small amount of mass actually reaches the NS, with the remainder being
expelled in some sort of outflow.  For convenience, we parametrize
$\dot M$ as a power-law in $r$
\begin{equation}
\dot M(r) = \left(\frac{r}{R}\right)^p \dot M_{\rm NS} ,
\label{eq:bb99}
\end{equation}
where $R$ is the NS radius, $M_{\rm NS}$ is the mass accretion rate
onto the NS, and $p>0$. (The $p=0$ case corresponds to the radially
uniform inflow rate that we just dismissed.)  This is the same
parameterization used in the ``ADIOS'' (adiabatic inflow-outflow
solution) model of \citet{bb99}. The electron density now varies as  
\begin{equation}
n_e(r) = \frac{2}{\mu\eta\sigma_T R(1+X)}\sqrt{\frac{c^2 R}{GM}}
      \left(\frac{\dot M_{\rm NS}}{\dot M_E}\right)
      \left(\frac{r}{R}\right)^{p-3/2}
\label{eq:n_er}
\end{equation}
and the optical depth is
\begin{eqnarray}
\tau_{\rm es} & = & \frac{2}{\mu\eta (1+X)}\sqrt{\frac{c^2 R}{GM}}
      \left(\frac{\dot M_{\rm NS}}{\dot M_E}\right) \nonumber \\
   && \times  \int_1^{r_c/R} \left(\frac{r}{R}\right)^{p-3/2} 
              d\left(\frac{r}{R}\right) \,.
\end{eqnarray}
For ionized gas with cosmic abundances, we have $\mu=0.6$ and $X=0.7$.
We require that $r_c<a$, where the binary separation is $a=2.8\times
10^{11}\,M_{1.9}^{1/3}$~cm \citep{fkr02,swd14}.  This is equivalent to
requiring that $r_c/R\lesssim 10^5$. Then taking $\dot M_{\rm NS}/\dot
M_E = 4\times 10^{-6}$ and $\tau_{\rm es}=10$, we find $1.4<p<1.6$ for
$0.1<\eta<1$.  Even larger $p$-values are required for smaller
coronas.  Such high $p$-values yield high electron densities far from
the NS, so that the scattering cloud would again itself become a source
of significant thermal bremsstrahlung emission.  As an example, we consider the
$p=1.5$ (uniform density) case, for which $n_e=4\times
10^{14}$~cm$^{-3}$.  The effective optical depth is $\tau_{\rm
  eff}\sim 10^{-4}$, so we expect optically thin bremsstrahlung
emission.  Taking $r_c/R=10^5$ and $kT_e=5$~keV, we
predict a bremsstrahlung emission measure $\int n_e^2 dV = 8\times
10^{62}$~cm$^{-3}$, which corresponds to a luminosity of $8.5\times
10^{39}$~erg~s$^{-1}$ $\approx 40 L_E$.  This is nearly 7 orders of
magnitude brighter than the observed quiescent emission!  We conclude
that a thermal Comptonization model is not physically self-consistent for the
hard spectral component in Cen~X-4 in quiescence and can be ruled out.
      
\subsection{Synchrotron shock emission}

Another way of producing a cut-off power-law spectrum is through
synchrotron emission. Radio pulsars dissipate their rotational energy
via a relativistic wind comprising charged particles and Poynting flux
\citep[see][and references therein]{aro02,gs06}.  Quiescent NS/LMXB
transients can turn on as radio pulsars under some circumstances
\citep{scc+94}; this has recently been observed in at least three
cases (see \S\ref{sec-comps}). If Cen X-4 turned on as a radio pulsar
during X-ray quiescence, then synchrotron shock emission from the
radio pulsar wind interacting with intrabinary material or the
interstellar medium (ISM) could produce a power-law X-ray spectrum
\citep{ccm+98}.  The synchrotron power-law spectrum will have a high-energy
exponential cutoff corresponding to the maximum energy of the electron
population accelerated in the shock.  This is generally thought to
occur at energies $\gtrsim$100~keV in pulsar wind shocks.  Indeed PSR 
J1023+0023, the only quiescent NS/LMXB in which synchrotron shock
emission is definitely thought to have been observed, has an unbroken
power-law spectrum out to at least 80~keV is observed
\citep{tya+14}. However, as we show below, it is possible to have 
lower cutoff energies in pulsar wind shocks. 

\subsubsection{Formalism}

We begin by reviewing the formalism for the synchrotron shock scenario
developed by \citet[hereinafter AT93]{at93} for the case of the
eclipsing ``black widow'' millisecond pulsar PSR B1957+20.  The pulsar
wind is powered by spin-down of the pulsar with an energy loss rate
set by magnetic dipole radiation,
\begin{equation}
  \dot E = \frac{4\pi^2 I \dot P}{P^3} ,
\end{equation}
where $P$ and $\dot P$ are the pulsar spin period and its
derivative\footnote{Note that $\dot P$ is the spin period derivative
  due to magnetic dipole 
  spin-down alone, in the absence of any accretion torques.}, and we assume
$I=10^{45}$~g~cm$^2$ for the NS moment of inertia.  Since we 
do not know $P$ and $\dot P$ for Cen X-4, we will scale the surface
dipole magnetic field as $B= 10^8\,B_8$~G and the spin period as
$P=3\, P_{\rm 3ms}$~ms, where we assume
\begin{eqnarray}
  B & = & \left(\frac{3c^3 I P\dot P}{8\pi^2 R^6}\right)^{1/2} \\
  & = & 1.8\times 10^8\, P_{\rm 3ms}^{1/2}\,
     \left(\frac{\dot P}{10^{-20}}\right)^{1/2} 
     \mbox{\rm\ G}\,.
\label{eq:pulsarB}
\end{eqnarray}
Thus, for a millisecond pulsar, we have $\dot P = 0.3\times
10^{-20}\,B_8^2\,P_{\rm 3ms}^{-1}$ and $\dot E = 4.8\times
10^{33}\,B_8^2\,P_{\rm 3ms}^{-3}$~erg~s$^{-1}$.

Upstream from the shock, the Lorentz factor in the wind is
\begin{equation}
\gamma_{\rm up} = 9\times 10^4\, \left(\frac{\eta_V}{0.3}\right)
       \left(\frac{Z}{A}\right) B_8\, P_{\rm 3ms}^{-3/2} ,
\end{equation}
where we assume that the ions in the wind have charge $Z$ and mass $A
m_p$ and are accelerated to a fraction $\eta_V\sim 0.3$ of the open
field line voltage of the NS.  Possible ion values range from protons
($Z=A=1$) to partially-ionized iron ($Z\sim 3$, $A=56$).  The relative
energetic contribution of Poynting flux and ions in the wind upstream
of the shock is described by a magnetization parameter \citep{kc84a},
\begin{equation}
\sigma = \frac{B_{\rm up}^2}{4\pi\rho_{\rm up}\gamma_{\rm up} c^2} ,
\end{equation}
where $B_{\rm up}$ and $\rho_{\rm up}$ are the magnetic field strength and density
of the upstream wind.  For a particle-dominated wind (like the one in
the Crab Nebula), $\sigma\sim 10^{-3}$, while for a magnetically
dominated wind, $\sigma\gg 1$.  The upstream magnetic field strength
can then be written as \citep{kc84b}
\begin{equation}
  B_{\rm up}(r_s) =  \left(\frac{\sigma}{1+\sigma}\right)^{1/2}
      \left(\frac{\dot E}{r_s^2\, c\, f_p}\right)^{1/2}
\label{eq:B1}
\end{equation}
where $f_p=\Delta\Omega_p/4\pi$ is the fractional solid angle into
which the wind is emitted.  From the shock jump conditions, the
downstream field strength is \citep{kc84a}
\begin{equation}
  B_{\rm down} \approx \left\{ \begin{array}{ll}
                3\,B_{\rm up} & \mbox{\ for $\sigma\ll 1$} \\
                B_{\rm up} & \mbox{\ for $\sigma\gg 1$} .
                \end{array} \right. 
\label{eq:B2}
\end{equation}

The shock will give rise to a power-law electron population with
energy distribution $N(\gamma)\propto \gamma^{-s}$ for
$\gamma>\gamma_{\rm up}$, with $s\sim 2$.  This will, in turn, produce a
synchrotron radiation spectrum with photon number index $\Gamma\sim
1.5$ for photon energies above
\begin{equation}
  E_{\rm min} \simeq 0.3\,\gamma_{\rm up}^2 
   \left(\frac{\hbar e B_{\rm down}}{m_e\, c}\right) ,
\label{eq:Emin}
\end{equation}
where $e$ and $m_e$ are the charge and mass of the electron.  
If radiative losses are negligible, these power laws will extend up to
a cutoff at Lorentz factor
\begin{equation}
\gamma_m =\left(\frac{A}{Z}\right)\left(\frac{m_p}{m_e}\right)\gamma_{\rm up} 
      = 2\times 10^8\,\left(\frac{\eta_V}{0.3}\right)\, 
      B_8\, P_{\rm 3ms}^{-3/2} ,
\end{equation}
and at photon energy
\begin{equation}
  E_c = \gamma_m^2
   \left(\frac{\hbar e B_{\rm down}}{m_e\, c}\right) .
\label{eq:Ecm}
\end{equation}
However, if radiative losses are significant, then the electron
population will extend only to $\gamma_s<\gamma_m$. To find
$\gamma_s$, we compare the acceleration time (AT93)
\begin{equation}
t_{\rm acc} = \left(\frac{A}{Z}\right)\frac{\gamma_{\rm up}\, m_p\, c}{e B_{\rm down}} ,
\end{equation}
and the synchrotron loss time
\begin{equation}
t_s(\gamma) = \frac{6\pi m_e\, c}{\sigma_T\, B_{\rm down}^2\,\gamma} ,
\end{equation}
and solve $t_{\rm acc} = t_s(\gamma_s)$. If $\gamma_s<\gamma_m$, then
radiative losses are important, and the photon power law will only
extend up to 
\begin{equation}
  E_c = \gamma_s^2 
   \left(\frac{\hbar e B_{\rm down}}{m_e\, c}\right) .
\label{eq:Ecs}
\end{equation}

Given the above formalism, we now consider two possible sites for the
pulsar wind shock location. 

\subsubsection{Synchrotron shock in the ISM}

If the shock occurs where the pulsar wind is confined by ram pressure
in the ISM, then the shock radius will be \citep[AT93]{kh88}
\begin{eqnarray}
r_s & = &  \left(\frac{\dot E}{4\pi f_p\, c\, m_p\, n\, v_p^2}\right)^{1/2} \\
 & = & 4\times 10^{15}\,f_p^{-1/2}\, n_1^{-1}\, v_{200}^{-2}\, 
        B_8\, P_{\rm 3ms}^{-3/2} \mbox{\rm\ cm} ,
\end{eqnarray}
where $n = n_1$~cm$^{-3}$ is the ISM particle density and $v_p =
200\,v_{200}$~km\,s$^{-1}$ is the pulsar space velocity \citep{swd14}.
At this large distance from the NS, we assume that $\sigma\ll 1$
\citep{aro02}.  Then, the upstream magnetic field strength is
\begin{equation}
  B_{\rm up}(r_s) = 3\times 10^{-6}\ \left(\frac{\sigma}{10^{-3}}\right)^{1/2} 
          n_1\, v_{200}^2 \mbox{\rm\ \ G} ,
\end{equation}
and the magnetic field downstream of the shock is $B_{\rm down}\approx 3B_{\rm up}$.
Radiative losses are negligible in such a weak field, so the expected
synchrotron power-law spectrum will extend from 
\begin{eqnarray}
 E_{\rm min} &=&  3\times 10^{-7}\, \left(\frac{\sigma}{10^{-3}}\right)^{1/2}
   \left(\frac{\eta_V}{0.3}\right)^2
   \left(\frac{Z}{A}\right)^2 \nonumber \\
  && \qquad \times\, n_1\, v_{200}^2\, B_8^2\, P_{\rm 3ms}^{-3} \mbox{\rm\ keV}
\end{eqnarray}
to a cutoff at 
\begin{eqnarray}
 E_c &= & 3.3\, \left(\frac{\sigma}{10^{-3}}\right)^{1/2}
    \left(\frac{\eta_V}{0.3}\right)^2\, 
   n_1\, v_{200}^2\nonumber \\
  && \qquad \times \, B_8^2\,  P_{\rm 3ms}^{-3} \mbox{\rm\ keV} .
\end{eqnarray}

\begin{figure}[t]
\begin{center}
\includegraphics[width=0.47\textwidth]{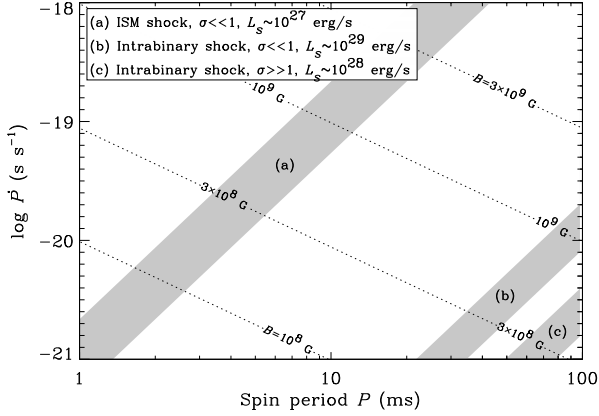}
\end{center}
\caption{Regions with low predicted cutoff energy $E_c$ for a
  synchrotron shock spectrum, displayed as a function of the radio
  pulsar spin period $P$ and period derivative $\dot P$.  $P$ and
  $\dot P$ are not known for Cen X-4.  The dotted lines show contours
  of constant pulsar magnetic field strength $B$ (see
  equation~[\ref{eq:pulsarB}]). The shaded regions have $E_c$ in the
  5--20~keV range, consistent with the observed power-law cutoff
  energy in Cen X-4, and correspond to the three cases listed in the
  box on the upper left. Radiative losses are negligible in all these
  regions ($\gamma_s>\gamma_m$). Contours of constant $E_c$ and
  constant synchrotron luminosity $L_s$ have the same slope on the
  $P$-$\dot P$ plane. In all three regions, the predicted $L_s$ is
  orders of magnitude too small to explain the observed power-law
  luminosity ($\sim 10^{32}$~erg~s$^{-1}$) in Cen X-4. Additional
  regions, corresponding to the intrabinary shock cases where
  radiative losses dominate ($\gamma_s<\gamma_m$), are not shown; they
  lie in an unphysical part of the phase space ($P\lesssim 1$~ms and
  $B\gtrsim 10^9$~G) for $E_c<20$~keV.  All calculations were done
  with $\eta_V=0.3$, $\epsilon_a=0.2$, and $f_{\rm geom}=0.05$.
  \label{fig_sync}}
\end{figure}

The ISM shock model produces a cutoff energy consistent with our
observed spectrum for a reasonable range of pulsar parameters, as
shown in shaded region (a) of Figure~\ref{fig_sync}. However, the
predicted synchrotron emission corresponding to that region is orders
of magnitude weaker than our observed power-law luminosity ($\sim
10^{32}$ erg~s$^{-1}$). Scaling to $P$ and $B$ values that lie in the
central strip of region (a) in Figure~\ref{fig_sync}, the
expected 0.3--20~keV synchrotron luminosity from the ISM shock is
(AT93, equation [16])  
\begin{eqnarray}
 L_s & \simeq & 2\times 10^{27}\ \left(\frac{\sigma}{10^{-3}}\right)
     \left(\frac{\epsilon_a}{0.2}\right)
     \left(\frac{\eta_V}{0.3}\right) 
     \left(\frac{n_1}{f_p^3}\right)^{1/2}\, v_{200} \nonumber \\
  && \quad \times \ \left(\frac{f_{\rm band}}{0.8}\right)\,
      B_8^4\, \left(\frac{P}{\mbox{\rm 2 ms}}\right)^{-6} 
       \mbox{\rm\ erg\,s$^{-1}$} ,
\label{eq-synclum_ism}
\end{eqnarray}
where $\epsilon_a$ is the conversion efficiency of pulsar wind
luminosity into particle acceleration in the shock, and $f_{\rm band}$
is the fraction of the bolometric synchrotron luminosity that lies in
the 0.3--20 keV band, 
\begin{equation}
 f_{\rm band} \simeq \left[\frac{\min(\mbox{20 keV},E_c)}{E_c}\right]^{1/2}
              - \left[\frac{\mbox{0.3 keV}}{E_c}\right]^{1/2} .
\end{equation}
Note that contours of constant $E_c$ and constant $L_s$ have the same
slope in Figure~\ref{fig_sync}, so the numerical pre-factor in
equation~(\ref{eq-synclum_ism})  characterizes the entire length of
the shaded region.  We conclude that synchrotron emission from a shock
in the ISM cannot explain the observed hard X-ray spectrum in Cen X-4. 

\subsubsection{Intrabinary shock}

Another possibility is that the pulsar wind collides with material
lost from the binary companion or the accretion flow at a shock
radius $r_s\sim a$, where the magnetic field is considerably stronger
than in the ISM case.  In PSR J1023+0038, the intrabinary shock was
modeled as occurring near the inner Lagrangian ($L_1$) point
\citep{bah+11}; in Cen X-4, the $L_1$ point lies at $r\approx 0.67 a$.
It is unclear whether the pulsar wind would be particle-dominated or
not in an intrabinary shock relatively close to the pulsar.
\citet{aro02} points out that theory predicts $\sigma\gg 1$ at the
pulsar light cylinder, but that observations indicate $\sigma\ll 1$ in
observed pulsar wind nebulae produced by ISM shocks, and that it is
not clear how or where the transition from one regime to the other
occurs. We note that \citet{bah+11} inferred $\sigma\gg1$ for the
intrabinary shock in PSR~J1023+0038.  We therefore consider both the
$\sigma\ll1$ and $\sigma\gg1$ cases here\footnote{We follow
  \citet{sgk+03} and \citet{bah+11} in taking this approach. However,
  it is not clear that the synchrotron shock model is viable for
  $\sigma\gg 1$. \citet{kc84a} wrote: ``Large-$\sigma$ shocks are
  effectively weak... Only when $\sigma\lesssim 0.1$ can a significant
  fraction of the total energy flux upstream be converted into thermal
  energy downstream and thereafter into synchrotron luminosity.''}.

Scaling to the binary separation in equation~(\ref{eq:B1}), the
upstream magnetic field is now 
\begin{eqnarray}
  B_{\rm up}(r_s) &=& \left(\frac{\sigma}{1+\sigma}\right)^{1/2}
    \left(\frac{\dot E}{a^2\,c\,f_p}\right)^{1/2} 
    \left(\frac{r_s}{a}\right)^{-1} \\
  &=&  0.04\ \left(\frac{\sigma}{10^{-3}}\right)^{1/2}
   f_p^{-1/2}\, \left(\frac{r_s}{a}\right)^{-1}\, 
   \left(\frac{a}{a_0}\right)^{-1}\, \nonumber \\
  && \qquad \times\ B_8\, P_{\rm 3ms}^{-3/2} \mbox{\rm\ G, for $\sigma\ll 1$}\\
 & = & 1.4\ f_p^{-1/2}\, \left(\frac{r_s}{a}\right)^{-1}\, 
     \left(\frac{a}{a_0}\right)^{-1}\, \nonumber \\
  && \qquad \times\ B_8\, P_{\rm 3ms}^{-3/2}\mbox{\rm\ G, for $\sigma\gg 1$} ,
\end{eqnarray}
where $a_0 = 2.8\times 10^{11}$~cm. The downstream field $B_{\rm
  down}$ is given by equation~(\ref{eq:B2}). From
equation~(\ref{eq:Emin}), the lower energy bound for the synchrotron
power law is
\begin{eqnarray}
 E_{\rm min} &=& 0.003\ \left(\frac{\sigma}{10^{-3}}\right)^{1/2}
       f_p^{-1/2}\,
       \left(\frac{\eta_V}{0.3}\right)^2
       \left(\frac{r_s}{a}\right)^{-1}
       \left(\frac{a}{a_0}\right)^{-1}  \nonumber \\
  && \qquad \times\ \left(\frac{Z}{A}\right)^2  
       \, B_8^3\, P_{\rm 3ms}^{-9/2}
       \mbox{\rm\ keV,\ \ for $\sigma\ll1$} \\
 &=& 0.04\ f_p^{-1/2}\,
       \left(\frac{\eta_V}{0.3}\right)^2
       \left(\frac{r_s}{a}\right)^{-1}
       \left(\frac{a}{a_0}\right)^{-1} \nonumber \\
  && \qquad \times\ \left(\frac{Z}{A}\right)^2 B_8^3\, P_{\rm 3ms}^{-9/2}
       \mbox{\rm\ keV,\ \ for $\sigma\gg1$} .
\end{eqnarray}
The upper bound depends on whether or not radiative losses are
important, which is set by the ratio
\begin{eqnarray}
\frac{\gamma_s}{\gamma_m} & = & 
     36\ \left(\frac{\sigma}{10^{-3}}\right)^{-1/2}
       f_p^{1/2}\, 
       \left(\frac{\eta_V}{0.3}\right)^{-2}
       \left(\frac{r_s}{a}\right) \, \left(\frac{a}{a_0}\right) \nonumber \\
   && \qquad  \times \ B_8^{-3}\, P_{\rm 3ms}^{9/2} ,
       \mbox{\rm\ \ for $\sigma\ll 1$} \\
 &=&     3\ f_p^{1/2}\, 
       \left(\frac{\eta_V}{0.3}\right)^{-2}
       \left(\frac{r_s}{a}\right) \, \left(\frac{a}{a_0}\right) \nonumber \\
   && \qquad  \times \ B_8^{-3}\, P_{\rm 3ms}^{9/2} ,
       \mbox{\rm\ \ for $\sigma\gg 1$} .
\end{eqnarray}
From equations~(\ref{eq:Ecm}) and (\ref{eq:Ecs}), the cutoff energy
for the $\sigma\ll 1$ case is then 
\begin{eqnarray}
 E_c &=& 4.5\times 10^4\ \left(\frac{\sigma}{10^{-3}}\right)^{1/2}
         f_p^{-1/2}
         \left(\frac{\eta_V}{0.3}\right)^2
         \left(\frac{r_s}{a}\right)^{-1} \nonumber \\
  && \quad \times\ \left(\frac{a}{a_0}\right)^{-1} B_8^3\, P_{\rm
           3ms}^{-9/2} 
       \mbox{\rm\ keV,\ \ for $\gamma_s>\gamma_m$} \\
  &=& 5.6\times 10^7\ \left(\frac{\sigma}{10^{-3}}\right)^{-1/2}
         f_p^{1/2}
         \left(\frac{\eta_V}{0.3}\right)^{-2}
         \left(\frac{r_s}{a}\right)\nonumber \\
  && \quad \times\ \left(\frac{a}{a_0}\right)\, B_8^{-3}\, P_{\rm 3ms}^{9/2}
       \mbox{\rm\ keV,\ \ for $\gamma_s<\gamma_m$} ,
\end{eqnarray}
and for the $\sigma\gg 1$ case is
\begin{eqnarray}
 E_c &=& 5.2\times 10^5\ f_p^{-1/2}
         \left(\frac{\eta_V}{0.3}\right)^2
         \left(\frac{r_s}{a}\right)^{-1} \, 
         \left(\frac{a}{a_0}\right)^{-1}\nonumber \\
  && \qquad \times\ B_8^3\, P_{\rm 3ms}^{-9/2}
       \mbox{\rm\ keV,\ \ for $\gamma_s>\gamma_m$} \\
  &=& 4.7\times 10^6\ f_p^{1/2}
         \left(\frac{\eta_V}{0.3}\right)^{-2}
         \left(\frac{r_s}{a}\right) \left(\frac{a}{a_0}\right) \nonumber \\
  && \qquad \times\ B_8^{-3}\, P_{\rm 3ms}^{9/2}
       \mbox{\rm\ keV,\ \ for $\gamma_s<\gamma_m$} .
\end{eqnarray}

The intrabinary shock model with $\gamma_s>\gamma_m$ is able to
produce a cutoff energy consistent with our observed spectrum for a
reasonable range of pulsar parameters, as shown in shaded regions (b)
and (c) of Figure~\ref{fig_sync}.  However, again, the predicted
synchrotron luminosity corresponding to those regions is orders of
magnitude smaller than our observed power-law luminosity. 
The expected 0.3--20~keV synchrotron luminosity from an intrabinary
shock is (AT93, equation [24])
\begin{equation}
 L_s  \simeq  \epsilon_a\, \dot E\, f_{\rm band}\, f_{\rm geom}
\end{equation}
where $f_{\rm geom}$ is the fraction of the pulsar wind intercepted by
intrabinary material. By analogy to PSR J1023+0038, we expect $f_{\rm
  geom}$ to be in the range 0.01--0.1 \citep{bah+11}.  Scaling to $P$
and $B$ values inside the shaded regions, we thus have
\begin{eqnarray}
 L_s & \simeq & 2\times 10^{29}\, \left(\frac{f_{\rm band}}{0.8}\right)\,
            \left(\frac{\epsilon_a}{0.2}\right)
            \left(\frac{f_{\rm geom}}{0.05}\right) \nonumber \\
   && \quad \times\ \left(\frac{B}{\mbox{$3\times 10^8$\ G}}\right)^2
            \left(\frac{P}{\mbox{40 ms}}\right)^{-3}
    \mbox{\rm\ erg\,s$^{-1}$},
\end{eqnarray}
for the $\gamma_s>\gamma_m$, $\sigma\ll 1$ case [region (b)], and 
\begin{eqnarray}
 L_s & \simeq & 3\times 10^{28}\, \left(\frac{f_{\rm band}}{0.8}\right)\,
            \left(\frac{\epsilon_a}{0.2}\right)
            \left(\frac{f_{\rm geom}}{0.05}\right) \nonumber \\
   && \quad \times\ \left(\frac{B}{\mbox{$3\times 10^8$\ G}}\right)^2
            \left(\frac{P}{\mbox{70 ms}}\right)^{-3}
    \mbox{\rm\ erg\,s$^{-1}$},
\end{eqnarray}
for the $\gamma_s>\gamma_m$, $\sigma\gg 1$ case [region (c)].  The
cases where radiative losses dominate ($\gamma_s<\gamma_m$) only give
low enough cutoff energies for an unphysical set of pulsar
parameters ($P\lesssim 1$~ms with $B\gtrsim 10^9$~G, beyond the upper
left corner of Figure~\ref{fig_sync}), and they predict a synchrotron
luminosity at least two orders of magnitude {\em larger} than what we
observe.  Taken together, we conclude that synchrotron emission from
an intrabinary shock cannot explain the cutoff power-law spectrum in
Cen X-4.   

\subsection{Bremsstrahlung}

A third way of modeling our cut-off spectrum is through thermal
bremsstrahlung emission from an optically thin cloud of hot electrons.
Our bremsstrahlung fits (see Tables~1 and 2) yielded an electron
temperature $kT_e=$18~keV and an emission measure
\begin{equation}
\int n_e^2\, dV = 1.6\times 10^{55}\, D_{\rm kpc}^2 {\rm\ cm}^{-3}\,.
\label{eq:bremss_em}
\end{equation}
We examine two different possibilities for the emission site.

\subsubsection{Emission from Above the NS Atmosphere}

We again consider the possibility of a hot, optically thin layer above
the NS atmosphere with geometric thickness $h= 10^3\,h_3$~cm. The
observed emission measure requires an electron density of
\begin{equation}
n_e= 3.6\times 10^{19}\,h_3^{-1/2}\,D_{\rm kpc}\mbox{\rm\ cm$^{-3}$}\,. 
\end{equation}
This implies a free-free absorption optical depth of only $\tau_{\rm
  ff}\simeq 10^{-8}$ at 2~keV, which is self-consistent for
optically-thin thermal bremsstrahlung emission.  In this scenario,
both the soft (thermal) emission and the hard (bremsstrahlung)
emission are formed in or above the NS atmosphere.  This can thus easily
account for the fact that the soft and hard emission are observed to
vary together on short time scales, with no detectable time lag. 

\citet{dds01} calculated the spectrum of an unmagnetized NS atmosphere
accreting at low rates through a RIAF flow. Their work differs from
the low-$\dot M$ case considered by \citet[see also \S3.2]{ztz+95} in
that it includes Coulomb heating of the atmosphere by energetic
protons in the RIAF flow.  Interestingly, they {\em predict} a hot,
optically thin surface layer above the NS atmosphere with $kT_e\approx
50$~keV and with significant bremsstrahlung emission expected for a
certain range of $\dot M_{\rm NS}$, in addition to the soft thermal
emission from the NS surface.  Their most detailed calculations were
made assuming a proton temperature of $0.5 k T_{\rm vir}$, where
$T_{\rm vir}$ is the virial temperature.  For this case, they found
that the bremsstrahlung emission is comparable in strength to the
thermal component when $\dot M_{\rm ns}\gtrsim 10^{-2} \dot M_E$, but
that it becomes negligible when $\dot M_{\rm ns}\lesssim 10^{-4} \dot
M_E$. By comparison, our observation of Cen X-4 measured comparable
luminosity in the the two components when the accretion rate was much
lower, $\dot M_{\rm ns}\leq 3.7\times 10^{-6} \dot M_E$. This is a
factor of $10^4$ discrepancy with the \citet{dds01} calculation.

One way of reconciling this is to invoke magnetic channeling of the
quiescent accretion flow onto the NS polar caps, thus 
increasing the local accretion rate per unit area.  However, this
would require very small polar caps with area $A_{\rm cap}\sim
10^{-4}\,\pi R^2$.  The measured spectral parameters of the
soft thermal component in Cen X-4 are not consistent with such small
polar caps (see Tables~1 and 2).  On the contrary, they suggest that
the thermal emission arises from a significant fraction of the stellar
surface.  Another alternative is to reexamine the assumptions of the
\citet{dds01} calculation, since the expected bremsstrahlung
luminosity must depend on the details of the Coulomb heating of the
atmosphere.  The authors found that, for fixed accretion rate,
reducing the proton temperature increases the heating in the upper
atmosphere, since the protons do not penetrate as deeply.  However, 
they did not investigate this case in detail, so it is not clear
whether a physically reasonable choice of proton temperature can
produce significant bremsstrahlung emission at accretion rates as
low as we observed in Cen X-4.  

We conclude that bremsstrahlung emission from a hot layer above the NS
atmosphere is consistent with our observed spectral cutoff.  However,
it is not yet clear how to produce sufficient luminosity to match our
data based on existing theoretical models for RIAF accretion onto
NSs. 

\subsubsection{Emission from the RIAF Accretion Flow}
\label{sec-bremss2}

As another alternative, we again assume that $\dot M$ varies with
distance from the NS according to equation~(\ref{eq:bb99}) and use
equation~(\ref{eq:n_er}) for $n_e$. Then, taking $\mu=0.6$ and $X=0.7$
for ionized gas with cosmic abundances, the expected emission measure
for a cloud of radius $r_c$ is
\begin{equation}
\int n_e^2\,dV  = 
 \frac{8\pi R}{p\eta^2\sigma_T^2}
 \left(\frac{c^2R}{GM}\right)
 \left(\frac{\dot M_{\rm NS}}{\dot M_E}\right)^2
 \left(\frac{r_c}{R}\right)^{2p} \,.
\end{equation}
Equating this to equation~(\ref{eq:bremss_em}), we find that
\begin{eqnarray}
\frac{r_c}{R} & \approx & 
  10^{4/p}\, p^{1/2p}\, D_{\rm kpc}^{1/p}\,
 \left(\frac{\eta}{0.1}\right)^{1/p} \nonumber \\
  & & \qquad \times
 \left(\frac{\dot M_{\rm NS}/\dot M_E}{4\times10^{-6}}\right)^{-1/p}
  ,
\end{eqnarray}
which sets the scale for the bremsstrahlung emission region in units
of the NS radius $R$.  Requiring $r_c<a$ or equivalently $r_c/R
\lesssim 10^5$, we find that $p\gtrsim 0.8$. Alternatively, if we
require that the cloud lies inside the RIAF transition radius
($r_c<r_t$ or $r_c/R\lesssim 10^4$), then we find that $p\gtrsim 1$.
In either case, only a small fraction $(r_c/R)^{-p}\lesssim 10^{-4}$
of the mass transferred reaches the NS in quiescence; the rest
accumulates at large $r$ or is expelled.  In particular, we note that
much of the bremsstrahlung emission is coming from electrons at large
$r$. For $p=1$, the electron density $n_{\rm e}$ ranges from
$10^{14}$~cm$^{-3}$ near the NS to $10^{12}$~cm$^{-3}$ near $r_c$.

This scenario is able to self-consistently account for the observed
bremsstrahlung luminosity.  However, it is challenging to explain 
the short ($\lesssim$30~s) lag time observed between the soft and hard
flux variability in the X-ray light curve (see \S3.1). Presumably, 
the hard flares would arise from mass fluctuations in the RIAF flow,
while the soft flares would arise when those fluctuations reach the NS
surface.  The shortest possible time scale between the soft and
hard flares is then the free-fall time scale from $r_c$, 
\begin{equation}
t_{\rm ff} = 60\, M_{1.9}^{-1/2}\, \left(\frac{r_c}{10^4\,R}\right)^{3/2}
     {\rm\ s}\,.
\end{equation}
This is only marginally consistent with our upper limit on the lag
time. We conclude that bremsstrahlung emission from the RIAF flow
is consistent with our observed spectral cutoff and luminosity, but
that placing this emission far from the NS is difficult to reconcile
with the fact that the soft and hard emission vary together on short
time scales.

\section{DISCUSSION}

\subsection{The spectral cutoff in Cen X-4}

We have measured a cutoff in the hard X-ray power-law spectrum of Cen
X-4 which can be fit with an exponential cutoff at around 10~keV or a
bremsstrahlung spectrum with $kT_e=18$~keV.  This is the first
detection of a power-law cutoff in this source class, and it finally
permits a more detailed investigation of the origin of the hard
component in quiescent NS/LMXBs.  We were able to rule out both thermal
Comptonization and synchrotron shock emission as the origin of the
spectral cutoff.  Instead, the hard X-ray spectrum can be understood as
bremsstrahlung emission, arising either from a hot, optically thin
corona above the NS atmosphere or from hot electrons in an optically
thin RIAF.  The NS atmosphere scenario has the advantage that it can
easily explain why the soft and hard emission varies together on short
time scales, while the RIAF scenario has the advantage that it can
self-consistently account for the observed luminosity. 


The 18~keV electron temperature for the {\tt bremss}
model is much lower than either the $\sim$50~keV electron temperature
predicted for the hot layer above a NS atmosphere \citep{dds01} or the
$\gtrsim$100~keV electron temperature expected in a RIAF flow around
a black hole \citep{mq97}.  This may be due to Compton cooling of the
bremsstrahlung electrons by the soft X-ray photons from the NS
atmosphere, in which case we would expect $T_e$ to depend upon the
soft X-ray luminosity $L_{\rm soft}$. The absence of a detectable
Compton emission component is not problematic. The Compton radiative
power density is 
\begin{equation}
P_C = \sigma_T\,n_e\,\left(\frac{4 kT_e}{m_e c^2}\right)
      \left(\frac{L_{\rm soft}}{4\pi r^2}\right) .
\label{eq:compt_radpow}
\end{equation}
The resulting inverse Compton luminosity is
\begin{eqnarray}
L_{C, {\rm atm}} & = & 5\times 10^{29}\,h_3^{1/2}\,
     \left(\frac{kT}{\mbox{\rm 18 keV}}\right) \nonumber \\
 & & \quad \times \left(\frac{L_{\rm soft}/L_E}{2\times 10^{-6}}\right)
     \mbox{\rm\ erg~s$^{-1}$}
\end{eqnarray}
for cooling above the NS atmosphere, or 
\begin{eqnarray}
L_{C, {\rm RIAF}} & = & 3\times 10^{30}\,
     \left(\frac{\eta}{0.1}\right)^{-1}
     \left(\frac{kT}{\mbox{\rm 18 keV}}\right)
   \left(\frac{\dot M_{\rm NS}/\dot M_E}{4\times 10^{-6}}\right)
     \nonumber \\
   & & \quad \times \left(\frac{L_{\rm soft}/L_E}{2\times 10^{-6}}\right)
     \mbox{\rm\ erg~s$^{-1}$}
\end{eqnarray}
for cooling in the accretion flow.  In either case, the Compton
luminosity is no more than a few percent of the overall source
luminosity, and thus essentially undetectable in our spectrum.

It is interesting to consider whether the flares observed in the X-ray
light curve might be expected to affect the electron temperature (and
thus the cutoff energy). The Compton cooling time scale is $t_C
=(3/2)n_e kT_e/P_C$. For emission above the NS atmosphere, this yields
\begin{equation}
t_{C,{\rm atm}}  =  1\times 10^{-2}\, M_{1.9}^{-1}\, 
   \left(\frac{L_{\rm soft}/L_E}{2\times 10^{-6}}\right)^{-1} 
        \mbox{\rm\ s} \,,
\end{equation}
so the cooling is nearly instantaneous.  For this scenario, we might
expect to measure spectral changes in the {\em NuSTAR} band
($\gtrsim$10~keV) during the flares, although we did not have
sufficient signal-to-noise to do this with our observation.  By
contrast, emission in the RIAF gives 
\begin{equation}
t_{C,{\rm RIAF}}  =  1\times 10^{6}\, M_{1.9}^{-1}\, 
   \left(\frac{r}{10^4 R}\right)^2
   \left(\frac{L_{\rm soft}/L_E}{2\times 10^{-6}}\right)^{-1} 
        \mbox{\rm\ s} \,,
\end{equation}
so that in this case short-term flaring behavior will not result in
significant Compton cooling of the bremsstrahlung electrons in
the RIAF flow.  This might provide an avenue for discriminating
between the two scenarios.

\begin{figure}[t]
\begin{center}
\includegraphics[width=0.47\textwidth]{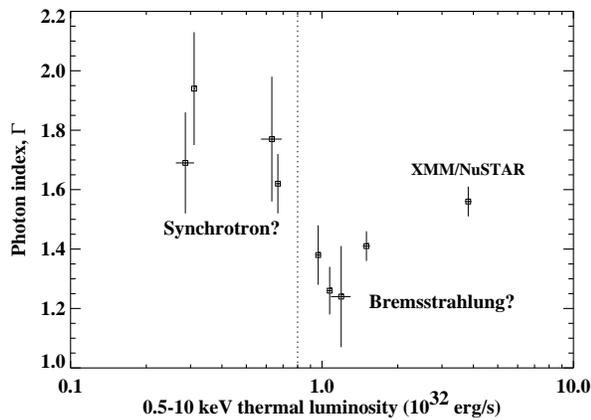}
\end{center}
\caption{Hard X-ray power-law photon index (2--10 keV) versus
  unabsorbed 0.5--10~keV thermal luminosity for deep observations of
  Cen X-4 (see Table~\ref{tbl-3}). The {\em XMM}/{\em NuSTAR} point is
  from a {\tt tbabs*(nsatmos+powerlaw)} fit to the 0.3--10~keV data
  only. The thermal luminosity $L_{\rm th}$ is computed for a distance
  of 1~kpc. For $L_{\rm th}\gtrsim 10^{32}\,D_{\rm kpc}^2$
  erg~s$^{-1}$, the data are roughly consistent with a trend of
  steeper power-law slope for higher $L_{\rm th}$ as expected for
  bremsstrahlung emission, if we assume that $kT_e$ is reduced from
  50--100~keV via Compton cooling by thermal photons.  In energy bandpasses
  well below $kT_e$, a bremsstrahlung spectrum is a $\Gamma=1$ power
  law.  The large $\Gamma$ measured at the lowest thermal luminosities
  may indicate a transition to synchrotron shock emission at extremely
  low $\dot M$.
  \label{fig_gamma}}
\end{figure}

On longer time scales (months to years), the fact that the slope of
the hard X-ray power-law spectrum in Cen X-4 was observed to vary from
epoch to epoch over the course of two decades \citep{cbm+10} makes it 
unlikely that the 18~keV bremsstrahlung spectrum we have measured is a
constant feature of the source in quiescence.  In fact, we can
demonstrate that the hard X-ray cutoff energy in Cen X-4 is likely
variable by noting that the departure of the hard X-ray spectrum from
an unbroken power-law above $\simeq$6~keV is evident in our 2013 {\em
  XMM-Newton} spectrum alone, even without including the {\em NuSTAR}
data. The shape of an 18~keV bremsstrahlung spectrum will show
noticeable curvature below 10~keV. On the other hand, all previous
deep observations of Cen X-4 in the 0.5--10~keV band are consistent
with an unbroken hard X-ray power-law spectrum, indicating a higher
cutoff energy for those observations. All of these observations
occurred at significantly lower luminosity (see Figure~\ref{fig_gamma}).  
We would expect a lower thermal luminosity to result in reduced
Compton cooling of the bremsstrahlung electrons and hence a higher
electron temperature, consistent with a higher cutoff energy.

\begin{deluxetable*}{llcccc}
\tablecaption{HARD X-RAY POWER-LAW SPECTRA OF DEEP CEN X-4 OBSERVATIONS
  \tablenotemark{a}\label{tbl-3}}  
\tablewidth{\textwidth}
\tablehead{ & & \colhead{Exposure} & \colhead{$L_{\rm th}$\tablenotemark{b}}&
   & \\ 
 \colhead{Start date} & \colhead{Mission} & \colhead{(ks)} &
 \colhead{($10^{32}$ erg s$^{-1}$)} & \colhead{$\Gamma$} &  \colhead{Ref.}
}
\startdata
1994 Feb 27 & {\em ASCA}   & 39  & 1.19(11) & 1.24(17) & 1 \\
2001 Aug 20 & {\em XMM}    & 53  & 1.50(5)  & 1.41(5)  & 1 \\
2003 Mar 1  & {\em XMM}    & 78  & 1.07(2)  & 1.26(8)  & 1 \\
2009 Jan 16 & {\em Suzaku} & 147 & 0.29(2)  & 1.69(17) & 1 \\
2010 Aug 25 & {\em XMM}    & 21  & 0.63(6)  & 1.77(21) & 2 \\
2010 Sep 4  & {\em XMM}    & 23  & 0.67(1)  & 1.62(10) & 2 \\
2011 Jan 24 & {\em XMM}    & 15  & 0.97(2)  & 1.38(10) & 2 \\
2011 Jan 31 & {\em XMM}    & 14  & 0.31(1)  & 1.94(19) & 2 \\
2013 Jan 20 & {\em XMM}+{\em NuSTAR}\tablenotemark{c} & 27/114 & 3.8(1) &
  1.56(5)  & 3 
\enddata
\tablenotetext{a}{1$\sigma$ uncertainties in last digits shown in
  parentheses.}
\tablenotetext{b}{0.5--10 keV thermal luminosity assuming $D$=1 kpc.}
\tablenotetext{c}{Fit only to 0.3--10 keV data using {\tt
    tbabs*(nsatmos+powerlaw)} model, with no power-law break.}
\tablerefs{(1) Cackett et al. 2010 and references therein; (2) Cackett
  et al. 2013; (3) This work.} 
\tablecomments{All archival data fit to {\tt phabs*(nsatmos+powerlaw)}
  model. {\em Chandra} observations excluded owing to possible photon pileup.}
\end{deluxetable*}

Moreover, we would expect the 2--10~keV power-law slope to vary
systematically with the thermal luminosity $L_{\rm th}$.  For energy
bandpasses far below $kT_e$, a bremsstrahlung spectrum is a $\Gamma=1$
power-law \citep[see, e.g.][]{rl79}; as one approaches the cutoff at
$kT_e$ from below, the effective $\Gamma$ over a fixed bandpass
increases as the spectrum begins to gradually roll over.  Thus, if we
assume that $kT_e\gtrsim$~50--100~keV in the absence of Compton
cooling, then a Compton-cooled bremsstrahlung model predicts that
$\Gamma$ should be close to 1 at low $L_{\rm th}$ and should increase
as $L_{\rm th}$ rises and $kT_e$ falls.  This relationship will
eventually break down when $kT_e$ gets sufficiently low, because an
unbroken power-law no longer provides even a rough fit to a sharp
spectral cutoff.  The archival data roughly support this picture for
$L_{\rm th}\gtrsim 10^{32}\,D_{\rm kpc}^2$ erg~s$^{-1}$. In
Figure~\ref{fig_gamma}, we plot $\Gamma$ versus $L_{\rm th}$ for deep
quiescent observations of Cen X-4 made since 1994, including our
observation\footnote{We exclude
  observations made with {\em Chandra}/ACIS, which may be subject to
  pileup effects. We note that the $\Gamma$ values found by
  \citet{cbm+10} for these observations are much smaller than for any
  other observations, as would be expected if there is significant
  pileup. We will reexamine these data elsewhere.}. These observations
are listed in Table~\ref{tbl-3}; the archival spectra are collected
from \citet{cbm+10} and \citet{cbd+13}. The trend of the observations
in Figure~\ref{fig_gamma} with $L_{\rm th}\gtrsim 10^{32}\,D_{\rm kpc}^2$ 
erg~s$^{-1}$ is roughly consistent with our expectation for a
Compton-cooled bremsstrahlung model.

At the lowest thermal lumnosities, however, $\Gamma$ jumps to higher
values. The abrupt change is suggestive of a spectral transition to a
different emission mechanism.  Given the extremely low luminosities, one
might consider coronal X-ray emission from the companion star
\citep{br00}, but the observed spectral shapes for these observations were not 
consistent with coronal emission \citep{cbm+10,cbd+13}.  Instead, we suggest
that this may indicate a transition to synchrotron shock emission at
the lowest luminosities.  We discuss this further in \S\ref{sec-comps}. 

\subsection{The Nature of Low-$\dot M$ Accretion}
\label{sec-lowMdot}

As in most quiescent transient NS/LMXBs, a basic requirement for Cen
X-4 is that most of the accretion flow does not reach the NS, since
the inferred $\dot M_{\rm NS}$ is substantially smaller than the
binary mass transfer rate $\dot M_T\sim 0.01 \dot M_E$
expected for a Roche-lobe--filing main sequence donor in a 15.1~hr
binary \cite[e.g.,][]{kkb96}.  There may be several mechanisms that
contribute to this.  First of all, the disk instability model for LMXB
transients predicts that most of the accretion flow during X-ray
quiescence builds up in the outer accretion disk until a thermal
instability ensues, causing an outburst \citep[see][]{las01}. At low
$\dot M$, the outer disk will transition into a quasi-spherical RIAF
flow at $r_t$.  It has previously been noted that RIAF models for quiescent NS
transients require that most of the RIAF flow is somehow prevented
from reaching the NS \citep{adh+98,men+99}. One way of achieving this
is the ADIOS-like outflow that we discussed in \S\ref{sec-bremss2}. 

Another possibility is that most of the flow reaching the NS
magnetosphere is centrifugally inhibited by the magnetic 
``propeller effect'' \citep{is75,ukr+06}.  This occurs when the
magnetosphere extends beyond the corotation radius (see
equation~[\ref{eq_corot}]) 
We define the magnetospheric radius $r_m$ as the location where the
magnetic and material stresses are equal,
\begin{eqnarray}
r_m & = & \xi \left(\frac{\mu_m^4}{GM\dot M^2}\right)^{1/7}  \\
 & = &  31\ \xi\,B_8^{4/7}\,M_{1.9}^{-1/7}\,R_{10}^{10/7} \nonumber \\
 & & \qquad \times\left(\frac{\dot M/\dot M_E}{0.01}\right)^{-2/7} 
             \mbox{\rm km} ,
\end{eqnarray}
where $\mu_m$ is the magnetic dipole moment of the NS, $\xi$ is an
order unity constant that depends upon the details of the accretion
flow near the magnetosphere \citep[see, e.g.,][]{pc99}, and the usual
$R^{12/7}$ scaling is modified by the $R$-dependence of $\dot M_E$.
In the propeller regime, $r_m>r_{\rm co}$. For ordinary thin-disk
magnetic accretion, the disk extends to the magnetosphere, and the
accretion is entirely shut off in this regime. 
However for a RIAF flow onto a millisecond pulsar, $r_m$ will
generally lie inside the transition radius $r_t$, so that the flow
onto the magnetosphere will be quasi-spherical.  In this geometry, a
small fraction of the flow is able to reach the NS despite the
centrifugal barrier present in the propeller regime \citep{men+99}.
Whether material is expelled in a strong or weak outflow, or else
accumulates outside $r_{\rm co}$ (e.g., a ``dead'' disk), depends upon
details of the disk-magnetosphere interaction
\citep{st93,ds10,ds12,lru+14}. However, observationally,
\citet{bcb+13} have shown that a strong propeller outflow can likely
be ruled out in Cen X-4. 

Our observations support the conclusion of \citet{bcb+13} that
low-level accretion is occurring during X-ray quiescence in Cen X-4,
indicating that a small fraction of the accretion flow must eventually
reach the NS.  However, it remains unclear what combination of the
above mechanisms ultimately controls what that fraction is. 

\subsection{Comparison to Other Low-$\dot M$ Systems}
\label{sec-comps}

After Cen X-4, the next brightest well-studied quiescent NS/LMXB
transient is Aql X-1.  Unlike Cen X-4, Aql X-1 has a relatively short
recurrence time scale of 1--2~yr.  There have been several recent
studies of its quiescent emission \citep{cfh+11,ccd+14,stn+14}, all
observing a soft thermal component and hard power-law component with
no cutoff. \citet{stn+14} argue that, for their 2007 {\em Suzaku}
observations ($L/L_E=$[3--9]$\times 10^{-5}\,M_{1.9}$ for an assumed
distance of 5.2~kpc), the most appropriate model for the hard
component is Comptonization in either an optically thin ($\tau_{\rm
  es}\simeq 0.3$), very hot ($kT_e>100$~keV) corona or an optically
thick ($\tau_{\rm es}>3$), somewhat cooler ($kT_e\sim 50$~keV) corona.
They do not find the same inconsistency between their measured
$\tau_{\rm es}$ and radially uniform accretion that we found in
equation~(\ref{eq_tau}).  This is a consequence of their $\dot M$
being higher and their $\tau_{\rm es}$ being lower than in our Cen X-4
observation. However, we noted in \S\ref{sec-compt} that $\tau_{\rm
  es}$ and $kT_e$ vary inversely, and the high $kT_e$ values (and
corresponding cutoff energies) that \citet{stn+14} fit lie above their
observation bandpass.  The {\em Suzaku} data are thus unable to rule
out the presence of a cutoff below 50--100~keV (but still above their
bandpass); this would introduce the same difficulties for a
Comptonization model that we found in Cen X-4, although it would be
somewhat mitigated by the higher $\dot M$. We note that our
bremsstrahlung model could explain the hard component in Aql~X-1 for
$kT_e\gtrsim 30$~keV.

It is interesting to also compare the behavior of Cen X-4 with systems that
have been observed to transition between LMXB and radio pulsar states
during X-ray quiescence.  The theoretical expectation is that such
transitions are controlled by location of the NS magnetospheric
boundary \citep{scc+94}.  We can compare $r_m$ to both $r_{\rm co}$
and the light-cylinder radius,
\begin{equation}
r_{\rm lc} = \frac{cP}{2\pi} = 144\,P_{\rm 3ms} \mbox{\rm\ km} \,.
\end{equation}
For sufficiently low $\dot M$, we have $r_{\rm co}<r_m<r_{\rm lc}$
and the system will be in the propeller regime, with accretion onto
the NS (mostly) shut off \citep{is75,ukr+06}.  For even lower $\dot
M$, we will have $r_{\rm co}<r_{\rm lc}<r_m$.  In this case, the radio
pulsar mechanism can turn on, with the radiation pressure of a radio
pulsar wind clearing the magnetosphere and an intrabinary shock giving
rise to synchrotron X-ray emission \citep{scc+94,ccm+98,bpd+01}. The
$\dot M$ implied for transition to a millisecond radio pulsar state
corresponds to $L_x\lesssim 10^{33}$~erg~s, where the exact value
depends on details of the system and the disk-magnetosphere
interaction. 

Indirect evidence for such transitions in NS/LMXBs during X-ray
quiescence was previously reported in Aql~X-1 \citep[$L_x=6\times
  10^{32}$ erg~s$^{-1}$;][]{csm+98} and SAX~J1808.4$-$3658
\citep[$L_x=5\times 10^{31}$][]{bdd+03,cdc+04,dht+08}.  However, more
direct evidence has been reported more recently in at least three
systems.  The 1.7~ms radio pulsar PSR J1023+0038 (hereafter J1023) has
made two transitions between the LMXB and radio pulsar states.  It is
now understood to have been in an LMXB state during 2000--2001, with
direct optical evidence for the presence of an accretion disk
\citep{wat+09}. However, a state transition then occurred, with
subsequent observations establishing the absence of an accretion disk
during 2002--2013 \citep{ta05} as well as the presence of a
millisecond eclipsing radio pulsar during 2007--2013 \citep{asr+09},
along with a low X-ray luminosity associated with intrabinary
synchrotron shock emission \citep{akb+10,bah+11}. A second state
transition was observed more recently, with the radio pulsar turning
off and an accretion disk reemerging \citep{sah+13,pah+14}.

In all of these observations, J1023 has remained in X-ray quiescence
in the sense that a high-luminosity ($L_x\gtrsim
10^{36}$~erg~s$^{-1}$) transient X-ray outburst was not
observed. However, two distinct sub-states are evident: a faint
X-ray--quiescent state ($L_x\sim 10^{32}$ erg~s$^{-1}$) during which
radio pulsations are seen, and a bright X-ray--quiescent state
($L_x\sim 10^{33}$ erg~s$^{-1}$) during which the radio pulsar is off.
This suggests that $r_m$ has moved outside the light cylinder in the
fainter state. In both sub-states, the 0.3--10~keV X-ray emission has
a power-law spectrum with little or no thermal component, with
$\Gamma=1.3$ in the faint state \citep{akb+10} and $\Gamma=1.69$ in
the  bright state \citep{pah+14}.  {\em NuSTAR}
observations show that these power-law spectra remain unbroken up to at least
79~keV \citep{tya+14}.  During faint X-ray--quiescence, the X-ray flux
shows high-amplitude modulation at the orbital period
\citep{akb+10,tya+14}, similar to what is seen in the X-ray emission
from some (but not all) eclipsing millisecond radio pulsars \citep[the
  so-called ``black widow'' and ``redback''
  systems;][]{rob12,rmg+14}. During bright X-ray--quiescence, the
X-ray flux shows strong, rapid flickering, with the intensity varying
by an order of magnitude on time scales $<$100~s
\citep{pah+14,tya+14}.

A second object in which two LMXB/radio pulsar state transitions were
seen is the M28 globular cluster source PSR J1824$-$2452I (hereafter
M28I), a 3.9~ms radio pulsar. The radio pulsar underwent a bright
transient X-ray outburst ($L_x\sim 10^{36}$ erg~s$^{-1}$) in 2013
March during which accretion-powered millisecond pulsations and a
thermonuclear X-ray burst were observed, establishing the system as an
NS/LMXB. The system returned to X-ray quiescence within a month, at which
point radio pulsations were again detected \citep{pfb+13}. These
observations demonstrate that LMXB/radio pulsar state transitions can
occur on time scales as short as days.  During X-ray quiescence, rapid
($\lesssim$500~s) intensity variations of nearly an order of magnitude
are again seen [(0.6--4)$\times 10^{33}$ erg~s$^{-1}$], with no change
in the 0.3--10~keV X-ray spectrum: an absorbed power-law with
$\Gamma=1.2$ and no detectable thermal component \citep{lbh+14}.
Remarkably, this rapid variability seems to toggle between two stable
flux levels; \citet{lbh+14} suggest that this represents fast
transitions between synchrotron shock emission and magnetospheric
accretion.  No orbital variability of the X-ray flux is reported in
M28I.

The third object in which an LMXB/radio pulsar transition was seen is
XSS~J12270$-$4859 (hereafter J12270), a faint hard X-ray source
associated with a relatively bright {\em Fermi} $\gamma$-ray source.
During 2003--2012, the source was in a quiescent LMXB state with an
absorbed power-law X-ray spectrum with $\Gamma=1.7$, no evidence for
a thermal spectral component, a 0.1--10~keV luminosity of $L_x=2\times
10^{33}\,D_{\rm kpc}$~erg~s$^{-1}$, highly variable X-ray flaring, and
multiwavelength evidence for the presence of an accretion disk
\citep{dfb+10,dbf+13}. In late 2012, the source made a transition to a
lower ($6\times 10^{31}\,D_{\rm kpc}$~erg~s$^{-1}$) luminosity state
with a power-law X-ray spectrum with $\Gamma=1.2$, a thermal fraction
$<$9\%, and a large-amplitude orbital modulation of the X-ray flux
\citep{bph+14,bpa+14}.  After this transition, 1.69~ms radio
pulsations were also detected \citep{rbr14}.

Cen X-4 has comparable luminosity to J1023, M28I, and J12270, and so
it is presumably at least close to the regime where LMXB/radio
transitions could occur. The rapid X-ray variability we observe (see
\S3.1) is quite different from the large-amplitude orbital modulation
seen in the low-quiescent state of J1023, but it is qualitatively
similar to (although somewhat weaker than) the flickering seen in the
high-quiescent states of J1023 and J12270 as well as the M28I
quiescent variability. On the other hand, unlike those three sources,
the X-ray spectrum of Cen X-4 has a substantial thermal fraction
($\simeq$60\%). This may indicate that more of the accretion flow
reaches the surface of Cen X-4 than in the other systems.  If we apply
our Cen X-4 Compton-cooled bremsstrahlung model to the hard X-ray
emission in J1023, M28I, and J12270 during their radio-quiet/X-ray
quiescent states, then we would expect a high (50--100~keV) electron
temperature and an unbroken 2--10~keV power-law X-ray spectrum,
consistent with what was observed.  

Of course, synchrotron shock emission can also produce an unbroken
power-law in the X-ray band, and we expect this mechanism to dominate
in the radio pulsar state.  At sufficiently low luminosity, Cen X-4
should transition into a radio pulsar state; we suggest that this may
be what occurs at $L_{\rm th}\lesssim 10^{32}\,D_{\rm kpc}^2$
erg~s$^{-1}$ in Figure~\ref{fig_gamma}, and that the jump in $\Gamma$
might reflect a sharp transition from high-temperature bremsstrahlung
emission to synchrotron shock emission.  This is consistent with the
suggestion by \citet{jgm+04} that the power-law component in quiescent
NS/LMXBs arises from accretion at higher $\dot M$ and from some a
different, non-accretion mechanism (e.g., synchroton shock emission)
at lower $\dot M$; they used this to explain how the fractional
power-law contribution to the quiescent luminosity varies with $\dot
M$ in quiescent NS/LMXBs.  It would be interesting to search for
millisecond radio pulsations from Cen X-4 when its X-ray luminosity
next drops to $\lesssim 10^{32}\,D_{\rm kpc}$~erg~s$^{-1}$.

\acknowledgments We thank the referee, Craig Heinke, for several
suggestions that greatly improved our paper. D.C. thanks Herman Marshall, Sera
Markoff, Caroline D'Angelo, Stephen Reynolds, Federico Bernardini, Phil
Charles, and Chris Done for useful discussions, and Luca Zampieri and
Roberto Soria for sharing their XSPEC additive table model {\tt
  zamp}. We also thank Thorsten Brand for help with evaluating the
level of photon pileup in the {\em XMM-Newton} data. This work was
supported in part under NASA contract NNG08FD60C and made use of data
from the {\it NuSTAR} mission, a project led by the California
Institute of Technology, managed by the Jet Propulsion Laboratory, and
funded by NASA. We thank the {\it NuSTAR} Operations, Software and
Calibration teams for support with the execution and analysis of these
observations.  This research has made use of the {\it NuSTAR} Data
Analysis Software (NuSTARDAS), jointly developed by the ASI Science
Data Center (ASDC, Italy) and the California Institute of Technology
(USA).  JAT acknowledges partial support from the {\em XMM-Newton}
Guest Observer program through NASA grant NNX13AB47G.

\medskip
{\em Facilities:} \facility{NuSTAR}, \facility{XMM}.

\end{document}